\documentclass[a4paper,11pt]{article}%
\pdfoutput=1
\usepackage{amssymb}
\usepackage{graphicx}
\usepackage{amsbsy}
\usepackage{amsmath}
\usepackage{cite}
\usepackage{slashed}
\usepackage{amsfonts}
\usepackage{hyperref}%
\hypersetup{
colorlinks   = true,
linkcolor    = blue,
citecolor    = blue,
}

\begin{document}
\thispagestyle{empty} \setcounter{page}{0} 
\begin{flushright}
March 2019\\
\end{flushright}

\vskip          4.1 true cm

\begin{center}
{\huge Axions are blind to anomalies}\\[1.9cm]

\textsc{J\'er\'emie Quevillon}$^{1}$\textsc{\ and Christopher Smith}$^{2}$%
\vspace{0.5cm}\\[9pt]\smallskip{\small \textsl{\textit{Laboratoire de Physique
Subatomique et de Cosmologie, }}}\linebreak%
{\small \textsl{\textit{Universit\'{e} Grenoble-Alpes, CNRS/IN2P3, Grenoble
INP, 38000 Grenoble, France}.}} \\[1.9cm]\textbf{Abstract}\smallskip
\end{center}

\begin{quote}
\noindent

The axion couplings to SM gauge bosons are derived in various models, and shown to always arise entirely from non-anomalous fermion loops. They are thus independent of the anomaly structure of the model. This fact is without consequence for vector gauge interactions like QCD and QED, but has a major impact for chiral gauge theories. For example, in the DFSZ axion model, the couplings of axions to electroweak gauge bosons do not follow the pattern expected from chiral anomalies, as we prove by an explicit calculation. The reason for this mismatch is traced back to triangle Feynman diagrams sensitive to the anomalous breaking of the vector Ward identity, and is ultimately related to the conservation of baryon and lepton numbers. Though our analyses are entirely done for true axion models, this observation could have important consequences for axion-like particle searches.

\let\thefootnote\relax\footnotetext{$^{1}\;$jeremie.quevillon@lpsc.in2p3.fr}%
\ \footnotetext{$^{2}\;$chsmith@lpsc.in2p3.fr}
\end{quote}

\newpage

\setcounter{tocdepth}{2}

\section{Introduction}

Axions have been around for more than 40 years, and even if the simplest models have been ruled out, they still remain probably the best solution for the strong $CP$ problem of the Standard Model. They come under many guises, but the basic principle is always the same: design a spontaneously broken global $U(1)_{PQ}$ symmetry and assign chiral charges to some colored fermions~\cite{PQ}. This ensures the associated Goldstone boson $a^{0}$, the axion~\cite{Weinberg:1977ma,Wilczek:1977pj}, has a direct coupling to gluons, and possibly also to photons, of the form%
\begin{equation}
\mathcal{L}_{eff}=\frac{a^{0}}{16\pi^{2}v}\left(  g_{s}^{2}\mathcal{N}_{C}G_{\mu\nu}^{a}\tilde{G}^{a,\mu\nu}+e^{2}\mathcal{N}_{em}F_{\mu\nu}\tilde{F}^{\mu\nu}\right)  \ ,\label{CouplingGG}%
\end{equation}
where $v$ is the vacuum expectation value (VEV) of the scalar breaking the $U(1)_{PQ}$ symmetry, and $\mathcal{N}_{C,em}$ some constants related to the fermion $U(1)_{PQ}$ and gauge charges. Naively, the $CP$-violating QCD coupling $\theta_{QCD}G_{\mu\nu}\tilde{G}^{\mu\nu}$ can then be rotated away no matter the value of $\theta_{QCD}$ since the Goldstone boson is shifted as $a^{0}\rightarrow a^{0}+v\theta$ under a $U(1)_{PQ}$ transformation of parameter $\theta$. In reality, what happens is that non-perturbative QCD effects dressing $G_{\mu\nu}\tilde{G}^{\mu\nu}$ create an effective potential for the axion field~\cite{PQ} (see also e.g. Ref.~\cite{diCortona:2015ldu}), whose minimum is attained precisely when $\left\langle a^{0}\right\rangle +v\theta_{QCD}=0$. In the process, the axion acquires a small QCD-induced mass, typically well below the eV scale~\cite{Bardeen:1977bd,Kim:1986ax}. For a recent review, see for example Ref.~\cite{Marsh:2015xka}.

Though this picture is correct, the nature of the couplings in Eq.~(\ref{CouplingGG}) is often wrongly ascribed to the anomaly of the $U(1)_{PQ}$ fermionic current. This idea comes from Noether's theorem: the Goldstone boson $a^{0}$ is coupled to its symmetry current, $vp^{\mu}=\langle0|J_{PQ}^{\mu}|a^{0}(p)\rangle$, and the current is anomalous, $\partial_{\mu}J_{PQ}^{\mu}\sim g_{s}^{2}\mathcal{N}_{C}G_{\mu\nu}^{a}\tilde{G}^{a,\mu\nu}+e^{2}\mathcal{N}_{em}F_{\mu\nu}\tilde{F}^{\mu\nu}$, hence it seems Eq.~(\ref{CouplingGG}) is immediately recovered. It has been known for a long time that this derivation of the axion couplings is not correct but sufficient for practical purpose~\cite{Georgi:1986df,Bardeen:1986yb}. It permits to identify the couplings of axions to gluons and photons as induced by heavy fermions. So, by a common abuse of language, the $a\rightarrow \gamma\gamma$ and $a\rightarrow gg$ processes are said to be induced by the anomaly in the $U(1)_{PQ}$ current.

However, as we will explicitly demonstrate in this paper, using this same procedure to derive the couplings of the axion to electroweak gauge bosons fails whenever chiral fermions are charged under $U(1)_{PQ}$. In that case, the true gauge couplings do not follow the pattern expected from the anomalies in the $U(1)_{PQ}$ current. More precisely, for the $SU(3)_{C}\otimes SU(2)_{L}\otimes U(1)_{Y}$ gauge group, one expects from chiral anomalies~\cite{Georgi:1986df} (see also Ref.~\cite{Alonso-Alvarez:2018irt})%
\begin{equation}
\mathcal{L}_{eff}=\frac{a^{0}}{16\pi^{2}v}\left(  g_{s}^{2}\mathcal{N}%
_{C}G_{\mu\nu}^{a}\tilde{G}^{a,\mu\nu}+g^{2}\mathcal{N}_{L}W_{\mu\nu}%
^{i}\tilde{W}^{i,\mu\nu}+g^{\prime2}\mathcal{N}_{Y}B_{\mu\nu}\tilde{B}^{\mu
\nu}\right)  \ ,\label{CouplingWW}%
\end{equation}
where $\mathcal{N}_{C,L,Y}$ are derived from the $U(1)_{PQ}$ charges and gauge quantum numbers of the SM fermions. While for QCD and QED, which are vector theories, the coefficients $\mathcal{N}_{C}$ and $\mathcal{N}_{em}=\mathcal{N}_{L}+\mathcal{N}_{Y}$ indeed tune the $a^{0}\rightarrow gg$ and $\gamma\gamma$ decays, we will prove that the $a^{0}\rightarrow\gamma Z$, $ZZ$, and $W^{+}W^{-}$ decay amplitudes are independent of $\mathcal{N}_{L}$ and $\mathcal{N}_{Y}$! The true processes driving these decays are not anomalous and their relative strengths do not follow the pattern expected from Eq.~(\ref{CouplingWW}). The reason for this mismatch between the anomalous couplings and the true axion couplings will be traced to the existence of additional anomalies, occurring only for chiral theories. These additional anomalies cancel exactly the contributions in Eq.~(\ref{CouplingWW}), leaving as remainder only non-local, non-anomalous processes.

The paper is organized as follow. To set the stage, we start in the next section by presenting in details an axion toy model~\cite{PQ}. Though simple, this model illustrates many important physical features of more realistic axion models. In particular, it will be clear that the axion couplings to gauge bosons are not anomalous, but that this has no quantitative consequence for photons or gluons. Then, before turning our attention to full-fledged axion models, we derive a number of important results for anomalies in Section~\ref{Ano}. Specifically, to be able to treat chiral gauge theories, the Ward identities applied to $AVV$ or $AAA$ triangle graphs have to be properly calculated, where $V_{\mu}=\bar{\psi}\gamma_{\mu}\psi$ and $A_{\mu}=\bar{\psi}\gamma_{\mu}\gamma_{5}\psi$ denote the vector and axial currents. Indeed, it is necessary to go beyond simple regularization procedures to be able to locate the anomaly of the $AVV$ triangle in one of the vector Ward identities or to break explicitly the Bose symmetry of the $AAA$ triangle. Equipped with these results, we turn to the Peccei-Quinn axion model in Section~\ref{PQaxion}, derive the correct axion couplings to gauge bosons, and identify precisely where the naive procedure leading to Eq.~(\ref{CouplingWW}) fails. This analysis is then trivially extended to invisible axion models. Finally, our results are summarized in Section~\ref{Ccl}, along with their implications for axion-like particle searches.

\section{An axionic toy model\label{Toy}}

Let us consider the following simple extension of QED. Starting with a massless Dirac fermion $\psi$, charged under a local symmetry $U(1)_{em}$, we add a scalar field $\phi$ charged under a $U(1)_{PQ}$ global symmetry. The complete Lagrangian is%
\begin{equation}
\mathcal{L}=-\frac{1}{4}F_{\mu\nu}F^{\mu\nu}+\bar{\psi}_{L}(i\slashed D)\psi_{L}+\bar{\psi}_{R}(i \slashed D)\psi_{R}+(y\phi\bar{\psi}_{L}\psi_{R}+h.c.)+\partial_{\mu}\phi^{\dagger}\partial^{\mu}\phi-V(\phi)\ ,\label{LagrSymm}%
\end{equation}
where $D^{\mu}=\partial^{\mu}+ieA^{\mu}$ and $e$ is the electric charge of $\psi$, and $\psi_{L,R}=1/2(1\mp \gamma_5)\psi$. When the Yukawa coupling is non-zero, $y\neq 0$, the fermion fields have to be charged under $U(1)_{PQ}$ also. If we normalize the scalar charge to $-1$, then under a $U(1)_{PQ}$ transformation of parameter $\theta$,%
\begin{equation}
\phi\rightarrow\exp(-i\theta)\phi\ ,\ \ \psi_{L}\rightarrow\exp(i\alpha
\theta)\psi_{L}\ ,\ \psi_{R}\rightarrow\exp(i(\alpha+1)\theta)\psi
_{R}\ .\label{ToyCharge}%
\end{equation}
At the classical level, this clearly leaves the Lagrangian invariant whatever the free parameter $\alpha$. Actually, this parameter corresponds to the conservation of the global fermion number, which is aligned with the electric charge in this simple one-fermion model.

When the global $U(1)_{PQ}$ symmetry is broken spontaneously, a Goldstone boson is left behind and will be identified with the axion~\cite{Weinberg:1977ma,Wilczek:1977pj}. So, we choose the potential as $V(\phi^{\dagger}\phi)=\mu^{2}\phi^{\dagger}\phi+\lambda(\phi^{\dagger}\phi)^{2}$ with $\mu^{2}<0$. Let us analyze the resulting theory using two different representations for the scalar field.

\subsection{Linear representation}

\begin{figure}[t]
\centering\includegraphics[width=0.70\textwidth]{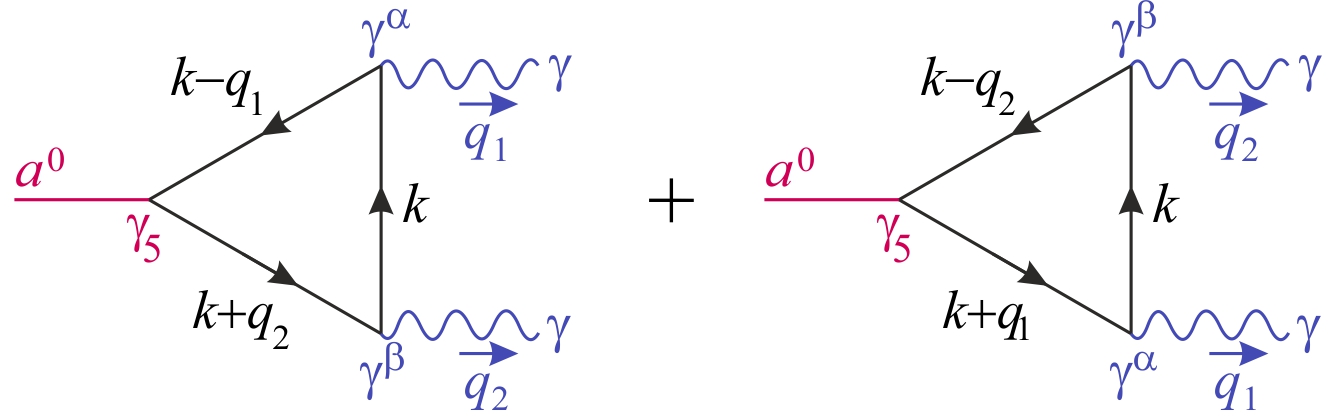}
\caption{Triangle graphs inducing the $a^{0}\rightarrow\gamma\gamma$ decay.}
\label{Fig1}
\end{figure}

Shifted to the true vacuum, the usual linear representation for the scalar field is
\begin{equation}
\phi(x)=\sigma^{0}(x)+ia^{0}(x)+v\ ,\label{LinearP}%
\end{equation}
with $v=-\mu^{2}/\lambda$. The Goldstone boson $a^{0}$ remains massless and is the axion, while the $\sigma^{0}$ field acquires a mass $m_{\sigma}^{2}=2\lambda v^{2}$. Assuming $y$ is real and denoting the fermion mass as $m\equiv vy$, the whole Lagrangian when $\phi$ is expended around the true vacuum becomes%
\begin{align}
\mathcal{L}_{\mathrm{Linear}}  & =-\frac{1}{4}F_{\mu\nu}F^{\mu\nu}+\bar{\psi
}(i\slashed D)\psi+m\bar{\psi}\psi\left(  1+\frac{\sigma^{0}}{v}\right)  +\frac{1}%
{2}\partial_{\mu}a^{0}\partial^{\mu}a^{0}+\frac{m}{v}a^{0}\bar{\psi}%
i\gamma_{5}\psi\nonumber\\
& +\frac{1}{2}(\partial_{\mu}\sigma^{0}\partial^{\mu}\sigma^{0}-2\lambda
v^{2}(\sigma^{0})^{2})-\lambda v\sigma^{0}((\sigma^{0})^{2}+(a^{0})^{2}%
)-\frac{\lambda}{4}((\sigma^{0})^{2}+(a^{0})^{2})^{2}-\frac{\lambda v^{4}}%
{4}\ .\label{LLagr}%
\end{align}
Since $\phi$ is not charged under $U(1)_{em}$, $a^{0}$ does not directly couple to photons. However, this coupling arises through one-loop triangle graphs, see Fig.~\ref{Fig1}. Adopting the Pauli-Villars procedure to regulate the loop amplitude in intermediate steps, we compute%
\begin{align}
\mathcal{T}_{PVV}^{\alpha\beta} &  =\int\frac{d^{4}k}{(2\pi)^{4}%
}(-1)\operatorname*{Tr}\left[  \frac{i}{\slashed k-\slashed q_{1}-m}\gamma^{\alpha
}\frac{i}{\slashed k-m}\gamma^{\beta}\frac{i}{\slashed k+\slashed q_{2}-m}\gamma
_{5}\right]  +(1,\alpha\leftrightarrow2,\beta)+(m\rightarrow M)\ \nonumber\\
&  =-i\frac{1}{2\pi^{2}}\varepsilon^{\alpha\beta\rho\sigma}q_{1,\rho
}q_{2,\sigma}(mC_{0}(m^{2})-MC_{0}(M^{2}))\ .
\label{TPVV}
\end{align}
The three-point scalar function $C_{0}(m^{2})\equiv C_{0}(q_{1}^{2},q_{2}^{2},(q_{1}+q_{2})^{2};m^{2},m^{2},m^{2})$ obeys\footnote{We use the notations and conventions of Ref.~\cite{Shtabovenko:2016sxi}} %
\begin{equation}
\underset{m\rightarrow\infty}{\lim}C_{0}(m^{2})=\frac{-1}{2m^{2}%
}\ ,\ \ \underset{m\rightarrow0}{\lim}m^{2}C_{0}(m^{2})=0\ .
\end{equation}
The regulator $M$ can thus safely be sent to infinity, and the decay amplitude in the linear representation for the scalar field is%
\begin{equation}
\mathcal{M}(a^{0}\rightarrow\gamma\gamma)_{\mathrm{Linear}}=-i\frac{m}{v}%
e^{2}\mathcal{T}_{PVV}^{\alpha\beta}\varepsilon(q_{1})_{\alpha}^{\ast
}\varepsilon(q_{2})_{\beta}^{\ast}=-\frac{e^{2}}{2\pi^{2}v}m^{2}C_{0}%
(m^{2})\varepsilon^{\alpha\beta\rho\sigma}\varepsilon(q_{1})_{\alpha}^{\ast
}\varepsilon(q_{2})_{\beta}^{\ast}q_{1,\rho}q_{2,\sigma}\ .\label{aggLin}%
\end{equation}
In the $m\rightarrow\infty$ limit, this amplitude corresponds to the local interaction%
\begin{equation}
\mathcal{L}_{\mathrm{Linear}}^{\mathrm{eff}}=-\frac{e^{2}}{16\pi^{2}v}%
a^{0}F_{\mu\nu}\tilde{F}^{\mu\nu}\ .\label{LinEff}%
\end{equation}
It is interesting to remark that this term is not decoupling \footnote{This non-decoupling is to be understood in the formal sense, since strictly
speaking the $m \rightarrow \infty $ limit is not compatible with our perturbative treatment of the Yukawa couplings.}, even though $\mathcal{T}_{PVV}^{\alpha\beta}\rightarrow0$ as $m\rightarrow\infty$, thanks to the $m$ factor coming from the $a^{0}$ coupling to the fermion. Still, at the level of the linearly-realized theory, anomalies do not show up and the loop amplitude can be safely computed using a naive regularization procedure, even if we know from the fermion charges in Eq.~(\ref{ToyCharge}) that $U(1)_{PQ}$ is anomalous.%

\subsection{Polar representation}

Instead of the linear representation for the scalar field, there is another representation more suited to the circular geometry of the vacuum\footnote{Note that the polar representation still allows the pseudoscalar of having a small mass which could be treated as a perturbation as this is usually done in axion phenomenology.}. The polar or exponential representation is 
\begin{equation}
\phi(x)=\frac{1}{\sqrt{2}}(v+\sigma^{0}(x))e^{-ia^{0}(x)/v}\ .\label{PolarP}%
\end{equation}
Then, under the action of a $U(1)_{PQ}$ rotation of parameter $\theta$, the component fields have simple (though non-linear) transformation properties: the $\sigma$ field is constant, $\sigma^{0}\rightarrow\sigma^{0}$, while the axion is shifted by a constant, $a^{0}\rightarrow a^{0}+v\theta$. Importantly, and even if we are using the same notations $a^{0}$ and $\sigma^{0}$, these fields are not the same ones as in Eq.~(\ref{LinearP}). In particular, the $a^{0}$ couples differently to fermions. Clearly, expanding the exponential, couplings of arbitrarily high dimensions are present, and the theory is no longer manifestly renormalizable. 

As a next step, after plugging the polar representation of $\phi$ in the Lagrangian, we perform a reparametrization of the fermion fields to make them invariant under $U(1)_{PQ}$:%
\begin{equation}
\psi_{L}(x)\rightarrow\exp(i\alpha a^{0}(x)/v)\psi_{L}(x)\ ,\ \psi
_{R}(x)\rightarrow\exp(i(\alpha+1)a^{0}(x)/v)\psi_{R}(x)\ .\label{Reparam}%
\end{equation}
This has three effects. First, the Goldstone field disappears from the Yukawa coupling. Second, the fermion kinetic terms induce derivative interactions,%
\begin{equation}
\delta\mathcal{L}_{\mathrm{Der}}=-\frac{\partial_{\mu}a^{0}}{v}(\alpha
\bar{\psi}_{L}\gamma^{\mu}\psi_{L}+(\alpha+1)\bar{\psi}_{R}\gamma^{\mu}%
\psi_{R})=-\frac{\partial_{\mu}a^{0}}{2v}((2\alpha+1)\bar{\psi}\gamma^{\mu
}\psi+\bar{\psi}\gamma^{\mu}\gamma_{5}\psi)\ ,
\end{equation}
where one can recognize the fermionic current associated to the $U(1)_{PQ}$ symmetry. Third, the fermionic path integral measure is not invariant, and the Jacobian of the transformation has to be included as a new local interaction:%
\begin{equation}
\delta\mathcal{L}_{\mathrm{Jac}}=\frac{e^{2}}{16\pi^{2}v}a^{0}(\alpha
-(\alpha+1))F_{\mu\nu}\tilde{F}^{\mu\nu}=-\frac{e^{2}}{16\pi^{2}v}a^{0}%
F_{\mu\nu}\tilde{F}^{\mu\nu}\ .\label{NLlocal}%
\end{equation}
The whole non-linear Lagrangian is thus%
\begin{align}
\mathcal{L}_{_{\mathrm{Polar}}} &  =-\frac{1}{4}F_{\mu\nu}F^{\mu\nu}+\bar
{\psi}(i \slashed D)\psi+m\bar{\psi}\psi\left(  1+\frac{\sigma^{0}}{v}\right) \nonumber\\
&  +\frac{1}{2}\partial_{\mu}a^{0}\partial^{\mu}a^{0}\left(  1+\frac
{\sigma^{0}}{v}\right)  ^{2}-\frac{\partial_{\mu}a^{0}}{2v}((2\alpha
+1)\bar{\psi}\gamma^{\mu}\psi+\bar{\psi}\gamma^{\mu}\gamma_{5}\psi
)-\frac{e^{2}}{16\pi^{2}v}a^{0}F_{\mu\nu}\tilde{F}^{\mu\nu}\nonumber\\
&  +\frac{1}{2}(\partial_{\mu}\sigma^{0}\partial^{\mu}\sigma^{0}-2\lambda
v^{2}(\sigma^{0})^{2})-\lambda v(\sigma^{0})^{3}-\frac{\lambda}{4}(\sigma
^{0})^{4}-\frac{\lambda v^{4}}{4}\ .\label{NLLagr}%
\end{align}
Under this form, the Lagrangian is manifestly $U(1)_{PQ}$-symmetric but for the local Jacobian term $a^{0}F_{\mu\nu}\tilde{F}^{\mu\nu}$ since $\sigma^{0}$, $\psi_{L,R}$, and $A^{\mu}$ are invariant but $a^{0}$ is shifted by a constant. This permits to get rid of any pre-existing $\theta_{em}F_{\mu\nu}\tilde{F}^{\mu\nu}$ term in the Lagrangian, by shifting $a^{0}\rightarrow a^{0}+v\theta_{em}$. Of course, for an abelian theory, $F_{\mu\nu}\tilde{F}^{\mu\nu}$ is a total derivative that can be safely discarded, but this toy model can easily be adapted to the non-abelian case, and closely resembles the KSVZ invisible axion model~\cite{KSVZ} when applied to QCD.

Let us now compute $a^{0}\rightarrow\gamma\gamma$ within the non-linearly realized theory. Together with the local amplitude from $\delta\mathcal{L}_{\mathrm{Jac}}$, we must include the triangle graphs arising from $\delta\mathcal{L}_{\mathrm{Der}}$, which have both a vector and axial component. The vector current $V^{\mu}=\bar{\psi}\gamma^{\mu}\psi$ does not contribute since the photon couplings are also vectorial, and the corresponding triangle graph vanishes thanks to Furry's theorem. The axial current $A^{\mu}=i\bar{\psi}\gamma^{\mu}\gamma_{5}\psi$, on the other hand, gives a non-zero contribution. We recognize the well-known $AVV$ triangle graph:%
\begin{equation}
\mathcal{T}_{AVV}^{\gamma\alpha\beta}=\int\frac{d^{4}k}{(2\pi)^{4}%
}(-1)\operatorname*{Tr}\left[  \frac{i}{\slashed k-\slashed q_{1}-m}\gamma^{\alpha
}\frac{i}{\slashed k-m}\gamma^{\beta}\frac{i}{\slashed k+\slashed q_{2}-m}\gamma^{\gamma
}\gamma_{5}\right]  +(1,\alpha\leftrightarrow2,\beta)\ .
\end{equation}
Adopting a Pauli-Villars regularization, we add to this a second term with $m\rightarrow M$, and proceed with the calculation in $d=4$ dimensions. With the derivative from $\partial_{\mu}a^{0}$ amounting to taking the divergence, we find%
\begin{align}
\mathcal{M}(a^{0}\overset{}{\rightarrow}\gamma\gamma)_{\mathrm{Der}}  &
=\frac{-e^{2}}{2v}i(q_{1}+q_{2})_{\gamma}\mathcal{T}_{AVV}^{\gamma\alpha\beta
}\varepsilon(q_{1})_{\alpha}^{\ast}\varepsilon(q_{2})_{\beta}^{\ast
}\nonumber\\
& =\frac{-e^{2}}{4\pi^{2}v}(2m^{2}C_{0}(m^{2})-2M^{2}C_{0}(M^{2}))\varepsilon
^{\alpha\beta\rho\sigma}\varepsilon(q_{1})_{\alpha}^{\ast}\varepsilon
(q_{2})_{\beta}^{\ast}q_{1,\rho}q_{2,\sigma}\ .\label{AxialDer}%
\end{align}
The regulator term does not decouple in the $M\rightarrow\infty$ limit but gives a finite local contribution. Interestingly, this contribution precisely cancel with the local term from $\delta\mathcal{L}_{\mathrm{Jac}}$, so altogether, we recover precisely the result obtained in the linear case, see Eq.~(\ref{aggLin}):%
\begin{align}
\mathcal{M}(a^{0}\overset{}{\rightarrow}\gamma\gamma)_{_{\mathrm{Polar}}} &
=\mathcal{M}(a^{0}\rightarrow\gamma\gamma)_{\mathrm{Der}}+\mathcal{M}%
(a^{0}\rightarrow\gamma\gamma)_{\mathrm{Jac}}\\
&  =-\frac{e^{2}}{2\pi^{2}v}m^{2}C_{0}(m^{2})\varepsilon^{\alpha\beta
\rho\sigma}\varepsilon(q_{1})_{\alpha}^{\ast}\varepsilon(q_{2})_{\beta}^{\ast
}q_{1,\rho}q_{2,\sigma}=\mathcal{M}(a^{0}\rightarrow\gamma\gamma
)_{\mathrm{Linear}}\ .
\end{align}

This exercise illustrates a number of important features:

\begin{itemize}
\item In the polar representation, the appearance of the anomalous local term $a^{0}F_{\mu\nu}\tilde{F}^{\mu\nu}$ is spurious. When computing specific amplitudes, it only serves to cancel out the anomalous term arising from the derivative interaction $\partial_{\mu}a^{0}\bar{\psi}\gamma^{\mu}\gamma
_{5}\psi$. At the end, this is nothing but an application of the well-known axial current Ward identity:
\begin{equation}
\partial_{\mu}A^{\mu}-\frac{e^{2}}{8\pi^{2}}F_{\mu\nu}\tilde{F}^{\mu\nu}=2imP\ ,\label{AxialWard}
\end{equation}
where $P=\bar{\psi}\gamma_{5}\psi$, $A^{\mu}=\bar{\psi}\gamma^{\mu}\gamma_{5}\psi$. The right-hand side corresponds to the amplitude in the linear representation, and the left-hand side to the two contributions arising in the polar representation. It is thus clear that it would be wrong to
understand the $a^{0}\rightarrow\gamma\gamma$ coupling as induced by the anomaly. The anomalous interaction of Eq.~(\ref{NLlocal}) only arises because of the reparametrization in Eq.~(\ref{Reparam}), and necessarily comes together with the appropriate derivative interactions.

\item In the $m\rightarrow0$ limit, the $a^{0}\rightarrow\gamma\gamma$ amplitude vanishes exactly,%
\begin{equation}
\mathcal{M}(a^{0}\rightarrow\gamma\gamma)_{\mathrm{Linear}}=\mathcal{M}(a^{0}\overset{}{\rightarrow}\gamma\gamma)_{_{\mathrm{Polar}}}\overset {m\rightarrow0}{=}0\ .
\end{equation}
In the linear representation, this trivially follows from the vanishing of the coupling of $a^{0}$ to fermions, see Eq.~(\ref{LLagr}). In the polar representation, it requires an exact cancellation of the local anomalous contribution with the non-local triangle amplitudes. Again, this can be understood in terms of the axial current Ward identity. In this limit, the spurious nature of the contact interaction is manifest.

\item The reason why $a^{0}\rightarrow\gamma\gamma$ is often misinterpreted as induced by the anomaly can be understood looking at the $m\rightarrow\infty$ limit. Indeed, the contribution of the derivative term vanishes,
\begin{equation}
\mathcal{M}(a^{0}\rightarrow\gamma\gamma)_{\mathrm{Der}}\overset
{m\rightarrow\infty}{=}0\ \ \ \ ,
\end{equation}
as can be trivially seen in Eq.~(\ref{AxialDer}). Since then all that remains in the non-linear theory is the local term from $\delta\mathcal{L}_{\mathrm{Jac}}$, it necessarily corresponds to the contribution surviving in the $m\rightarrow\infty$ limit in the linear theory, Eq.~(\ref{LinEff}). Yet, this is only a parametric correspondence, and certainly not a physical identification. The anomaly still cancels in $\mathcal{M}(a^{0}\rightarrow\gamma\gamma)_{\mathrm{Der}}+\mathcal{M}%
(a^{0}\rightarrow\gamma\gamma)_{\mathrm{Jac}}$, and the only surviving contribution is actually the first term of Eq.~(\ref{AxialDer}). Ultimately, the local anomalous term $\delta\mathcal{L}_{\mathrm{Jac}}$ in Eq.~(\ref{NLlocal}) is no more than a convenient book-keeping device tracking all the fields that have been integrated out~\cite{Georgi:1986df,Bardeen:1986yb}.

\item Furry's theorem together with the vector current Ward identity ensures the absence of any dependence on the free parameter $\alpha$ in the non-linear theory, in agreement with the absence of this parameter in the corresponding linear representation.

\item As an aside, it should be clear that though the anomalous coupling in Eq.~(\ref{NLlocal}) is not affected by radiative corrections~\cite{Adler:1969er}, neither is the anomalous part of the triangle graph in Eq.~(\ref{AxialDer}). So, the two exactly cancel at all orders, but the $a^{0}\rightarrow\gamma\gamma$ amplitude does get corrected by higher order effects since it is actually not induced by the anomaly. Further, note that the theory in Eq.~(\ref{LagrSymm}) is obviously renormalizable, so radiative corrections can be calculated perturbatively using standard techniques. In this respect, if the photons are also coupled to some other fermions $\chi$, the two-loop process $a^{0}\rightarrow\gamma\gamma
\rightarrow\bar{\chi}\chi$ is even UV finite in the linear representation~\cite{Ametller:1983ec}. This fact would clearly be difficult to guess using the polar representation, where the local $a^{0}\rightarrow\gamma\gamma$ vertex from $\delta\mathcal{L}_{\mathrm{Jac}}$ leads to a UV\ divergent diagram.

\item As said earlier, the present toy model can be generalized to more complicated, and more realistic KSVZ-like axion models~\cite{KSVZ}. Consider for instance the SM, to which a gauge-singlet scalar $\phi$ and a set of vector-like fermions $\psi$ are added:
\begin{equation}
\mathcal{L}_{\mathrm{KSVZ}}=\mathcal{L}_{\mathrm{SM}}+\bar{\psi}_{L}(i\slashed D)\psi_{L}+\bar{\psi}_{R}(i\slashed D)\psi_{R}+(y\phi\bar{\psi}_{L}\psi_{R}+h.c.)+\partial_{\mu}\phi^{\dagger }\partial^{\mu}\phi-V(\phi)\ ,\label{LKSVZ}%
\end{equation}
with the same scalar potential as in Eq.~(\ref{LagrSymm}). A priori, the covariant derivative can include all three SM interactions,
\begin{equation}
D^{\mu}=\partial^{\mu}-ig_{s}G_{a}^{\mu}T^{a}-igW_{i}^{\mu}T^{i}-ig^{\prime
}\frac{Y}{2}B^{\mu}\;,
\end{equation}
where $T^{a}$ and $T^{i}$ are the $SU(3)_{C}$ and $SU(2)_{L}$ generators in the representation carried by $\psi$, and $Y$ its hypercharge. Then, if $yv \gg v_{EW}$ and $v\gg v_{EW}$, with $v_{EW}$ the electroweak vacuum expectation value, the effect of the heavy fermion is to induce (see Eq.~(\ref{LinEff}))
\begin{equation}
\mathcal{L}_{\mathrm{Linear}}^{\mathrm{eff,KSVZ}}=-\frac{1}{16\pi^{2}v}%
a^{0}(g_{s}^{2}d_{L}^{\psi}C_{C}^{\psi}G_{\mu\nu}^{a}\tilde{G}^{a,\mu\nu
}+g^{2}d_{C}^{\psi}C_{L}^{\psi}W_{\mu\nu}^{i}\tilde{W}^{i,\mu\nu}+g^{\prime
2}d_{L}^{\psi}d_{C}^{\psi}C_{Y}^{\psi}B_{\mu\nu}\tilde{B}^{\mu\nu
})\ ,\label{KSVZlike}%
\end{equation}
with the quadratic invariants $C_{C}^{\psi}\delta^{ab}=\operatorname*{Tr}[T^{a}T^{b}]$, $C_{L}^{\psi}\delta^{ij}=\operatorname*{Tr}[T^{i}T^{j}]$, and $C_{Y}^{\psi}=Y^{2}/4$, and $d_{C,L}^{\psi}$ the corresponding dimensions of the $SU(3)_{C}$ and $SU(2)_{L}$ representations. Yet, several points must be clear: (1) These couplings are not anomalous, but result from the appropriately generalized Eq.~(\ref{aggLin}) in the $y\rightarrow\infty$ limit. (2) The integrated fermion must have vector couplings to gauge fields, otherwise the free parameter $\alpha$ corresponding to fermion number does not cancel in Eq.~(\ref{NLlocal}). We will see later on how to deal with chiral theories. (3) Even if vector-like, no mass term is allowed for $\psi$ because it would explicitly break the $U(1)_{PQ}$ symmetry. (4) The couplings in $\mathcal{L}_{\mathrm{Linear}}^{\mathrm{eff,KSVZ}}$ are not protected from radiative corrections since they are not anomalous. (5) Finally, the nature of the divergence arising when using this effective Lagrangian to compute e.g. the couplings of $a^{0}$ to SM fermions is clear. In the UV complete theory, Eq.~(\ref{LKSVZ}), the $a^{0}\rightarrow\gamma\gamma$, $WW$, and $gg$ vertices are never local. Instead, the pseudoscalar triangle of Eq.~(\ref{aggLin}) acts as a form factor and is sufficient to regulate the UV behavior of the vector boson loops, making them finite.
\end{itemize}

\section{On the consistent use of anomalies\label{Ano}}

In the course of the previous section, the anomaly appeared several times, and there is a number of peculiarities that are worth detailing. First, consider the calculation of the local term $\delta\mathcal{L}_{\mathrm{Jac}}$, see Eq.~(\ref{NLlocal}). The simplest way to compute the Jacobian of the transformation is to compute a triangle graph, this time with three left-(or right-)handed currents (the momentum flow is defined as in Fig.~\ref{Fig1}):%
\begin{equation}
\mathcal{T}_{LLL}^{\alpha\beta\gamma}=\int\frac{d^{4}k}{(2\pi)^{4}%
}(-1)\operatorname*{Tr}\left[  \frac{i}{\slashed k-\slashed q_{1}}\gamma^{\beta}%
P_{L}\frac{i}{\slashed k}\gamma^{\gamma}P_{L}\frac{i}{\slashed k+\slashed q_{2}}%
\gamma^{\alpha}P_{L}\right]  +(1,\beta\leftrightarrow2,\gamma)\ ,
\end{equation}
with $P_{L}=(1-\gamma_{5})/2$. Following Ref.~\cite{Weinberg:1996kr}, the divergences of this amplitude are calculated by first keeping track of the ambiguities in the loop momentum routing, and then from the surviving surface terms, and one finds:
\begin{subequations}
\label{AnoLLL}%
\begin{align}
i(q_{1}+q_{2})_{\alpha}\mathcal{T}_{LLL}^{\alpha\beta\gamma} &  =\frac{1}%
{8\pi^{2}}\left(  a-b\right)  \varepsilon^{\beta\gamma\mu\nu}q_{1\mu}q_{2\nu
}\ ,\\
-i(q_{1})_{\beta}\mathcal{T}_{LLL}^{\alpha\beta\gamma} &  =\frac{1}{8\pi^{2}%
}\left(  1+b\right)  \varepsilon^{\gamma\alpha\mu\nu}q_{1\mu}q_{2\nu}\ ,\\
-i(q_{2})_{\gamma}\mathcal{T}_{LLL}^{\alpha\beta\gamma} &  =\frac{1}{8\pi^{2}%
}\left(  1-a\right)  \varepsilon^{\alpha\beta\mu\nu}q_{1\mu}q_{2\nu}\ ,
\end{align}
where $a$ and $b$ are free parameters, such that no choice can force all three divergences to vanish simultaneously. For the purpose of deriving the $\delta\mathcal{L}_{\mathrm{Jac}}$ term (Eq.~(\ref{NLlocal})), since two currents have to be conserved to preserve the $U(1)_{em}$ invariance (see Eq.~(\ref{LagrSymm})), we can choose $a=-b=1$. Eq.~(\ref{NLlocal}) is then recovered including both $\psi_{L}$ and $\psi_{R}$, accounting for the fact that the anomalous terms are opposite for $\mathcal{T}_{LLL}^{\alpha\beta\gamma}$ and $\mathcal{T}_{RRR}^{\alpha\beta\gamma}$.

Yet, it is worth stressing that this choice $a=-b=1$ is not the one usually adopted, as it breaks the Bose symmetry of the amplitude under the exchange of the three currents. The consistent anomaly is that with $a=-b=1/3$, which restores the Bose symmetry as
\end{subequations}
\begin{equation}
i(q_{1}+q_{2})_{\alpha}\mathcal{T}_{LLL}^{\alpha\beta\gamma}|_{Bose}%
=-i(q_{1})_{\alpha}\mathcal{T}_{LLL}^{\gamma\alpha\beta}|_{Bose}=-i(q_{2})_{\alpha
}\mathcal{T}_{LLL}^{\beta\gamma\alpha}|_{Bose}=\frac{1}{12\pi^{2}}\varepsilon
^{\beta\gamma\mu\nu}q_{1\mu}q_{2\nu}\ .
\end{equation}
Importantly, this choice is not always explicit. For example, using the Pauli-Villars regularization, one immediately arrives at the Bose symmetric result (see Ref.~\cite{Bilal:2008qx}).

In terms of vector and axial currents, the $VVV$ and $VAA$ triangle amplitudes cancel out because of Furry's theorem. This leaves the $AVV$ and $AAA$ diagrams, for which we obtain using the same method as in Ref.~\cite{Weinberg:1996kr}:
\begin{subequations}
\label{AnoAVV0}%
\begin{align}
i(q_{1}+q_{2})_{\alpha}\mathcal{T}_{AVV}^{\alpha\beta\gamma}|_{m=0}\overset
{}{=}i(q_{1}+q_{2})_{\alpha}\mathcal{T}_{AAA}^{\alpha\beta\gamma}|_{m=0} &
=\frac{1}{4\pi^{2}}\left(  a-b\right)  \varepsilon^{\beta\gamma\mu\nu}q_{1\mu
}q_{2\nu}\ ,\\
-i(q_{1})_{\beta}\mathcal{T}_{AVV}^{\alpha\beta\gamma}|_{m=0}\overset{}%
{=}-i(q_{1})_{\beta}\mathcal{T}_{AAA}^{\alpha\beta\gamma}|_{m=0} &  =\frac
{1}{4\pi^{2}}\left(  1+b\right)  \varepsilon^{\gamma\alpha\mu\nu}q_{1\mu
}q_{2\nu}\ ,\\
-i(q_{2})_{\gamma}\mathcal{T}_{AVV}^{\alpha\beta\gamma}|_{m=0}\overset{}%
{=}-i(q_{2})_{\gamma}\mathcal{T}_{AAA}^{\alpha\beta\gamma}|_{m=0} &  =\frac
{1}{4\pi^{2}}\left(  1-a\right)  \varepsilon^{\alpha\beta\mu\nu}q_{1\mu
}q_{2\nu}\ .
\end{align}
\end{subequations}
These results are valid in the massless limit for the fermions. As for the consistent anomaly, the usual conventions are $a=-b=1$ for the $AVV$ triangle to keep the vector currents conserved, and $a=-b=1/3$ for the $AAA$ triangle to enforce the Bose symmetry. The former case is the basis for Eq.~(\ref{AxialDer}). These conventions are automatically enforced when using simple regularization procedures like Pauli-Villars or dimensional regularization. However, for the following, the freedom to assign the anomaly in one of the vector currents or to explicitly break the Bose symmetry will be essential, so these regularization procedures are inadequate.

At the level of the $AVV$ and $AAA$ triangles, there is no obstruction to consider massive fermions. These contributions can be separated from the anomalous pieces, and do not depend on the chosen momentum routing. Actually, a simple calculation leads to

\begin{subequations}
\label{AnoAVVm}%
\begin{align}
i(q_{1}+q_{2})_{\alpha}\mathcal{T}_{AVV}^{\alpha\beta\gamma} &
=2im\mathcal{T}_{PVV}^{\beta\gamma}+\frac{1}{4\pi^{2}}\left(  a-b\right)
\varepsilon^{\beta\gamma\mu\nu}q_{1\mu}q_{2\nu}\ ,\\
-i(q_{1})_{\beta}\mathcal{T}_{AVV}^{\alpha\beta\gamma} &  =\frac{1}{4\pi^{2}%
}\left(  1+b\right)  \varepsilon^{\gamma\alpha\mu\nu}q_{1\mu}q_{2\nu}\ ,\\
-i(q_{2})_{\gamma}\mathcal{T}_{AVV}^{\alpha\beta\gamma} &  =\frac{1}{4\pi^{2}%
}\left(  1-a\right)  \varepsilon^{\alpha\beta\mu\nu}q_{1\mu}q_{2\nu}\ ,
\end{align}
\end{subequations}
and
\begin{subequations}
\label{AnoAAAm}%
\begin{align}
i(q_{1}+q_{2})_{\alpha}\mathcal{T}_{AAA}^{\alpha\beta\gamma} &
=2im\mathcal{T}_{PAA}^{\beta\gamma}+\frac{1}{4\pi^{2}}\left(  a-b\right)
\varepsilon^{\beta\gamma\mu\nu}q_{1\mu}q_{2\nu}\ ,\\
-i(q_{1})_{\beta}\mathcal{T}_{AAA}^{\alpha\beta\gamma} &  =2im\mathcal{T}%
_{PAA}^{\alpha\gamma}+\frac{1}{4\pi^{2}}\left(  1+b\right)  \varepsilon
^{\gamma\alpha\mu\nu}q_{1\mu}q_{2\nu}\ ,\\
-i(q_{2})_{\gamma}\mathcal{T}_{AAA}^{\alpha\beta\gamma} &  =2im\mathcal{T}%
_{PAA}^{\alpha\beta}+\frac{1}{4\pi^{2}}\left(  1-a\right)  \varepsilon
^{\alpha\beta\mu\nu}q_{1\mu}q_{2\nu}\ ,
\end{align}
\end{subequations}
with the pseudoscalar triangle amplitudes (see Eq.~(\ref{TPVV}))
\begin{subequations}
\label{TriPVVPAA}%
\begin{align}
\mathcal{T}_{PVV}^{\alpha\beta}  & =-i\frac{1}{2\pi^{2}}mC_{0}(m^{2}%
)\varepsilon^{\alpha\beta\mu\nu}q_{1\mu}q_{2\nu}\ ,\\
\mathcal{T}_{PAA}^{\alpha\beta}  & =-i\frac{1}{2\pi^{2}}m(C_{0}(m^{2}%
)+2C_{1}(m^{2}))\varepsilon^{\alpha\beta\mu\nu}q_{1\mu}q_{2\nu}\ ,
\end{align}
\end{subequations}
where $C_{i}(m^{2})=C_{i}(q_{1}^{2},q_{2}^{2},(q_{1}+q_{2})^{2},m^{2},m^{2},m^{2})$.

\begin{figure}[t]
\centering\includegraphics[width=0.75\textwidth]{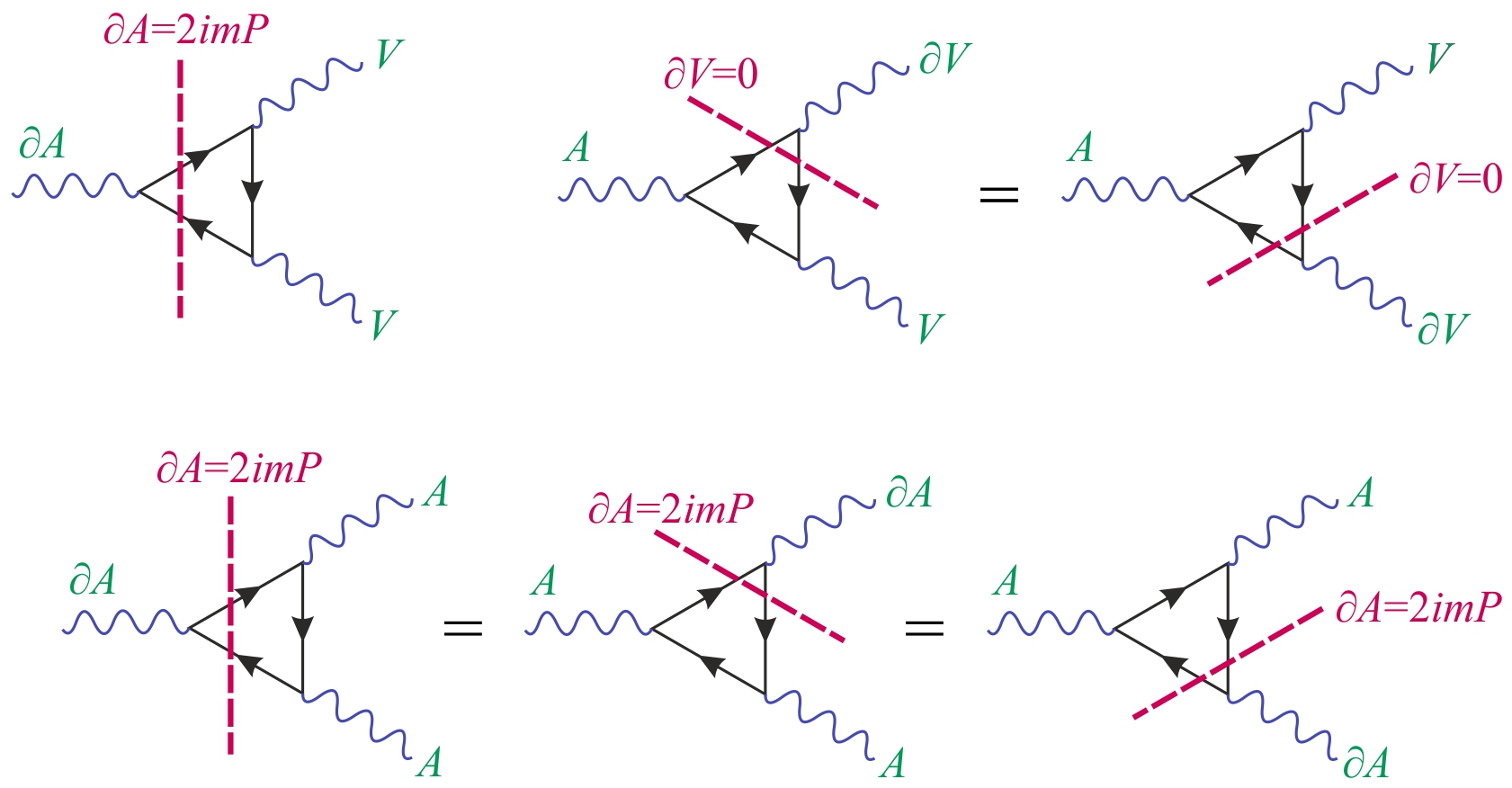}
\caption{Non--anomalous Ward identities valid for the mass-dependent terms of the triangle amplitudes. The dashed lines denote the cuts along which the fermionic equation of motion is valid.}%
\label{Fig2}
\end{figure}

In some sense, the usual conventions are automatic for the mass terms: the vector currents are always conserved and Bose symmetry must be manifest for the $AAA$ triangle. Indeed, once the anomalous terms are separated, the rest obeys the naive Ward identities $\partial_{\mu}V^{\mu}=0$ and $\partial_{\mu}A^{\mu}=2imP$, with $P$ the pseudoscalar current. This means that the mass-dependent term must be absent from $-i(q_{1})_{\beta}\mathcal{T}_{AVV}^{\alpha\beta\gamma}$ and $-i(q_{2})_{\gamma}\mathcal{T}_{AVV}^{\alpha\beta\gamma}$, and it must be equal for all three divergences of $\mathcal{T}_{AAA}^{\alpha\beta\gamma}$. This picture is trivially confirmed using dispersion relations~\cite{Frishman:1980dq}, see Fig.~\ref{Fig2}: cutting out a vector current necessarily gives zero since the intermediate fermions are on their mass shell and $\partial_{\alpha}\bar{\psi}\gamma^{\alpha}\psi=0$, while cutting out any one of the axial currents, the identity $\partial_{\alpha}\bar{\psi}\gamma^{\alpha}\gamma_{5}\psi=2im\bar{\psi}\gamma_{5}\psi$ must hold.%

Finally, in the $m\rightarrow\infty$ limit, using $C_{0}\rightarrow -1/(2m^{2})$ and $C_{1}(m^{2})\rightarrow1/(6m^{2})$, the various divergences are
\begin{subequations}
\label{AnoAVVinf}%
\begin{align}
i(q_{1}+q_{2})_{\alpha}\mathcal{T}_{AVV}^{\alpha\beta\gamma}|_{m\rightarrow
\infty} &  =\frac{1}{4\pi^{2}}\left(  a-b-2\right)  \varepsilon^{\beta
\gamma\mu\nu}q_{1\mu}q_{2\nu}\ ,\\
-i(q_{1})_{\beta}\mathcal{T}_{AVV}^{\alpha\beta\gamma}|_{m\rightarrow\infty}
&  =\frac{1}{4\pi^{2}}\left(  1+b\right)  \varepsilon^{\gamma\alpha\mu\nu
}q_{1\mu}q_{2\nu}\ ,\\
-i(q_{2})_{\gamma}\mathcal{T}_{AVV}^{\alpha\beta\gamma}|_{m\rightarrow\infty}
&  =\frac{1}{4\pi^{2}}\left(  1-a\right)  \varepsilon^{\alpha\beta\mu\nu
}q_{1\mu}q_{2\nu}\ ,
\end{align}
\end{subequations}
and
\begin{subequations}
\label{AnoAAAinf}%
\begin{align}
i(q_{1}+q_{2})_{\alpha}\mathcal{T}_{AAA}^{\alpha\beta\gamma}|_{m\rightarrow
\infty} &  =\frac{1}{4\pi^{2}}\left(  a-b-\frac{2}{3}\right)  \varepsilon
^{\beta\gamma\mu\nu}q_{1\mu}q_{2\nu}\ ,\\
-i(q_{1})_{\beta}\mathcal{T}_{AAA}^{\alpha\beta\gamma}|_{m\rightarrow\infty}
&  =\frac{1}{4\pi^{2}}\left(  1+b-\frac{2}{3}\right)  \varepsilon
^{\gamma\alpha\mu\nu}q_{1\mu}q_{2\nu}\ ,\\
-i(q_{2})_{\gamma}\mathcal{T}_{AAA}^{\alpha\beta\gamma}|_{m\rightarrow\infty}
&  =\frac{1}{4\pi^{2}}\left(  1-a-\frac{2}{3}\right)  \varepsilon^{\alpha
\beta\mu\nu}q_{1\mu}q_{2\nu}\ .
\end{align}
\end{subequations}
Crucially for the following, the vanishing of all the divergences in Eq.~(\ref{AnoAVVinf}) in the $m\rightarrow\infty$ limit is only true when the anomaly is carried by the axial current in the $AVV$ triangle, which implies the parametric choice $a=-b=1$. Similarly, the vanishing of all the divergences in Eq.~(\ref{AnoAAAinf}) in the $m\rightarrow\infty$ limit is only true when the anomaly is carried equally by the three axial currents in the $AAA$ triangle, which implies the choice $a=-b=1/3$. But evidently, there is no reason for this to hold in general and we will see in the next section how this plays a role in the derivation of the axion couplings to electroweak gauge bosons.

\section{The Peccei-Quinn axion and its couplings}
\label{PQaxion}

Historically, the first axion model was proposed by Peccei and Quinn~\cite{PQ}. It is based on a Two Higgs Doublet Model (THDM), but with an enlarged symmetry. Denoting the Higgs doublets as

\begin{equation}
\Phi_{1}=\left( \begin{array}[c]{c} H_{1}^{+}\\ H_{1}^{0} \end{array} \right)  \;,\;
\Phi_{2}=\left( \begin{array}[c]{c} H_{2}^{+}\\ H_{2}^{0} \end{array}\right)  \;,
\end{equation}
the scalar potential is assumed to allow for the independent rephasing%
\begin{equation}
U(1)_{1}\otimes U(1)_{2}: \Phi_{j}\rightarrow\exp(i\theta_{j})\Phi_{j}\ \quad \text{with} \ j =1,2.
\label{U1ch}
\end{equation}
In the standard notation (see Ref.~\cite{Gunion:1989we}), this forces the parameters $m_{12}$, $\lambda_{5,6,7}$ to vanish. One combination of these $U(1)$s can be identified with the hypercharge, since both doublets have the same gauge quantum numbers, while another combination is denoted $U(1)_{PQ}$. The vacuum expectation values%
\begin{equation}
\langle\Phi_{1}\rangle=\frac{1}{\sqrt{2}}\left( \begin{array}[c]{c} 0\\ v_{1} \end{array} \right)  \;,\;\langle\Phi_{2}\rangle=\frac{1}{\sqrt{2}}\left( \begin{array}[c]{c} 0\\ v_{2} \end{array} \right)  \ ,\label{THDMvev}%
\end{equation}
break $SU(2)_{L}\otimes U(1)_{Y}\otimes U(1)_{PQ}\rightarrow U(1)_{em}$ spontaneously, where $v_{1}^{2}+v_{2}^{2}\equiv v^{2}\approx\left(246\,\text{GeV}\right)  ^{2}$ and $\tan\beta\equiv v_{1}/v_{2}$. The $U(1)_{Y}$ symmetry is local and its Goldstone boson is eaten up by the $Z$ boson, but the $U(1)_{PQ}$ is only global and the massless left over state is the axion. Indeed, assuming Yukawa couplings of Type II,
\begin{equation}
\mathcal{L}_{\text{Yukawa}}=-\bar{u}_{R}\mathbf{Y}_{u}q_{L}\Phi_{1}-\bar
{d}_{R}\mathbf{Y}_{d}q_{L}\Phi_{2}^{\dagger}-\bar{e}_{R}\mathbf{Y}_{e}\ell
_{L}\Phi_{2}^{\dagger}+h.c.\;,\label{YukQuark}%
\end{equation}
the $U(1)_{PQ}$ symmetry is necessarily chiral and anomalous. Note that at this level, there is a lot of freedom in how to assign $U(1)_{1}\otimes U(1)_{2}$ charges for the fermions, and we will see later on how to (partially) fix them. Remark also that we did not include a phase difference between $v_{1}$ and $v_{2}$ in Eq.~(\ref{THDMvev}) because it can be removed by a $U(1)_{PQ}$ transformation. Actually, the $U(1)_{PQ}$ symmetry being anomalous, this rather moves the phase difference between $v_{1}$ and $v_{2}$ into the strong and electroweak $\theta$ terms. This is precisely the interplay that permits to solve the strong $CP$ puzzle once the $U(1)_{PQ}$ symmetry is spontaneously broken, since it entangles the phase difference with the axion field.

To further explore this simple model, we follow the same strategy as for the toy model of Section~\ref{Toy}, and perform the analysis using either a linear or a polar representation for the scalar fields $\Phi_{1,2}$.

\subsection{Linear representation}

The most convenient basis for the scalar states is obtained by performing a rotation of angle $\beta$,%
\begin{equation}
\left( \begin{array}[c]{c}
\Phi_{h}\\
\Phi_{H}%
\end{array} \right)  
=\left( \begin{array}[c]{cc}
\cos\beta & \sin\beta\\
-\sin\beta & \cos\beta
\end{array} \right)  
\left( \begin{array}[c]{c}
\Phi_{2}\\
\Phi_{1}%
\end{array}\right)  \ ,\label{RotBeta}%
\end{equation}
so that the VEV is carried by one doublet only%
\begin{equation}
\Phi_{h}=\frac{1}{\sqrt{2}}\exp\{i \boldsymbol{\sigma} \cdot \boldsymbol{G}(x)/v\}\left(
\begin{array}[c]{c}
0\\
v+\phi_{h}%
\end{array}
\right)  \;,\;\;\Phi_{H}=\frac{1}{\sqrt{2}}\left(
\begin{array}[c]{c}
\sqrt{2}H^{+}\\
\phi_{H}+iA^{0}%
\end{array}
\right)  \ ,\label{HiggsBasis}%
\end{equation}
where only the SM Goldstone bosons have been exponentiated. The pseudoscalar $A^{0}$ is the massless Goldstone boson corresponding to the $U(1)_{PQ}$ breaking, and it is parametrized linearly. The scalar states $(\phi_{h},\phi_{H})$ are related to the mass eigenstates $(h^{0},H^{0})$ through a rotation of angle $\alpha$ which depends on the scalar potential parameters, and $H^{+}$ is the charged scalar boson. We note in passing that the charge assignments corresponding to $U(1)_{1}\otimes U(1)_{2}\leftrightarrow U(1)_{Y}\otimes U(1)_{PQ}$ thus require orthogonality of the Goldstone states $G^{0}$ and $A^{0}$, and cannot be guessed a priori from the $U(1)$ charges of the doublets in Eq.~(\ref{U1ch}). Actually, since $A^{0}$ is a $\beta$-dependent combination of $\operatorname{Im}H_{1}^{0}$ and $\operatorname{Im}H_{2}^{0}$, the $U(1)_{PQ}$ charges of the doublets are also function of $\beta$. We will come back to this point later, and for now, we simply proceed by plugging Eq.~(\ref{HiggsBasis}) in the THDM Lagrangian.

Because $\Phi_{H}$ has no vacuum expectation value, the $A^{0}$ boson does not couple to pairs of gauge bosons at tree level. On the other hand, it does couple to fermions. From Eq.~(\ref{YukQuark}), we derive

\begin{equation}
\mathcal{L}_{A^{0}f\bar{f}}=-i\sum_{f=u,d,e}\frac{m_{f}}{v}\chi_{P}^{f}\,A^{0}%
\bar{\psi}_{f}\gamma_{5}\psi_{f}\ ,\ \ \chi_{P}^{u}=\frac{1}{\tan\beta
}\ ,\ \chi_{P}^{d}=\chi_{P}^{e}=\tan\beta\ .
\end{equation}
To reach this form, the mass terms are identified as $\sin\beta v\mathbf{Y}_{u}\equiv\sqrt{2}\mathbf{m}_{u}$ and $\cos\beta v\mathbf{Y}_{d,e}\equiv \sqrt{2}\mathbf{m}_{d,e}$ and the fermions are rotated to their mass basis. At the one-loop level, the $A^{0}$ then couples to pairs of gauge bosons through the $PVV$ and $PAA$ triangle graphs. Denoting the fermion currents coupled to gauge bosons as
$\bar{\psi}_{f^{\prime}}(g_{V}^{f}\gamma^{\mu}-g_{A}^{f}\gamma^{\mu}\gamma_{5})\psi_{f}$ with, using the short-hand $c_{W}=\cos\theta_{W},s_{W}=\sin\theta_{W}$:%
\begin{equation}%
\begin{tabular}
[c]{lllll}\hline
& $g$ & $W$ & $\gamma$ & $Z$\\\hline
$g_{V}^{f}\ =\ \ $ & $g_{s}T_{a}^{f}\ \ \ ,\ \ $ & $\dfrac{g}{\sqrt{2}}%
T_{3}^{f}\ \ ,\ \ $ & $eQ^{f}\ \ \ ,\ \ $ & $\dfrac{g}{2c_{W}}(T_{3}%
^{f}-2s_{W}^{2}Q^{f})\ ,$\\
$g_{A}^{f}\ =\ \ $ & $0\ \ \ ,\ \ $ & $\dfrac{g}{\sqrt{2}}T_{3}^{f}\ \ ,\ $ &
$0\ \ \ ,\ \ $ & $\dfrac{g}{2c_{W}}T_{3}^{f}\ .$\\\hline
\end{tabular}
\end{equation}
where $SU(3)$ generators $T_{a}^{f}$ are normalized such that $\operatorname*{Tr}(T_{a}^{f}T_{b}^{f})=1/2\delta^{ab}$, it is a simple exercise to find (see Eq.~(\ref{TriPVVPAA})):%
\begin{equation}
\mathcal{M}(A^{0}\rightarrow V_{1}V_{2})_{\mathrm{Linear}}=-i\sigma%
\sum_{f=u,d,e}\frac{m_{f}}{v}\chi_{P}^{f}\left(  g_{V,V_1}^{f}g_{V,V_2}%
^{f^{\prime}}\mathcal{T}_{PVV}^{\alpha\beta}(m_{f})+g_{A,V_1}^{f}g_{A,V_2}%
^{f^{\prime}}\mathcal{T}_{PAA}^{\alpha\beta}(m_{f})\right)  \varepsilon
(q_{1})_{\alpha}^{\ast}\varepsilon(q_{2})_{\beta}^{\ast}\ ,\label{LinAmpli}%
\end{equation}
where $\varepsilon(q_{1})_{\alpha}^{\ast}$ and $\varepsilon(q_{2})_{\beta}^{\ast}$ are the polarizations of $V_{1}$ and $V_{2}$, carrying respectively the momentum $q_{1}$ and $q_{2}$. To account for the absence of crossed diagrams for $W^{+}W^{-}$, we also define $\sigma=1/2$ for $V_{1,2}=W$, $\sigma=1$ for $V_{1,2}=g,\gamma,Z$. Also, the trace over color is understood for quarks, bringing a factor $\delta^{ab}/2$ for $V_{1,2}=g$, and $N_{C}$ for $V_{1,2}=W,Z,\gamma$. Finally, the loop functions in $\mathcal{T}_{PVV}^{\alpha\beta}(m_{f})$ and $\mathcal{T}_{AVV}^{\alpha\beta}(m_{f})$ are the same as in Eq.~(\ref{TriPVVPAA}) for $V_{1,2}=g,\gamma,Z$, and generalized as $C_{i}(m_{f}^{2})=C_{i}(q_{1}^{2},q_{2}^{2},(q_{1}+q_{2})^{2},m_{f}^{2},m_{f^{\prime}}^{2},m_{f}^{2}) $ with $f^{\prime}$ the $SU(2)$ partner of $f$ for $W^{+}W^{-}$. In that case, the CKM entries sum up to one when the three generations are included. The expression Eq.~(\ref{LinAmpli}) for the decay amplitudes agrees with the standard result of Ref.~\cite{Gunion:1991cw}.

A few limiting cases are interesting. First, remark that%
\begin{equation}
\left.  \mathcal{M}(A^{0}\rightarrow V_{1}V_{2})_{\mathrm{Linear}}\right\vert
_{m_{f}\rightarrow0}=0\ ,
\end{equation}
since the $A^{0}$ coupling to fermions is proportional to their masses. Second, in the opposite limit of infinite fermion masses, the loop functions behave as $C_{0}\rightarrow-1/(2m^{2})$ and $C_{1}(m^{2})\rightarrow 1/(6m^{2})$, so%
\begin{equation}
m_{f}\mathcal{T}_{PVV}^{\alpha\beta}(m_{f})=i\frac{1}{4\pi^{2}}\varepsilon
^{\alpha\beta\mu\nu}q_{1\mu}q_{2\nu}\ ,\ \ m_{f}\mathcal{T}_{PAA}^{\alpha
\beta}=i\frac{1}{12\pi^{2}}\varepsilon^{\alpha\beta\mu\nu}q_{1\mu}q_{2\nu}\ .
\end{equation}
The only special case is that of the lepton contribution to $A^0\rightarrow W^{+}W^{-}$, since neutrinos remain massless. In that case, $C_{0}(m_{e}^{2})\rightarrow-1/m_{e}^{2}$ and $C_{1}(m_{e}^{2})\rightarrow1/(2m_{e}^{2})$ in the $m_{e}\rightarrow\infty$ limit, and $\mathcal{T}_{PAA}^{\alpha\beta}$ cancels out. Plugging all these limits in the amplitudes, they take the simple form%
\begin{equation}
\left.  \mathcal{M}(A^{0}\overset{}{\rightarrow}V_{1}V_{2})_{\mathrm{Linear}%
}\right\vert _{m_{u,d,e}\rightarrow\infty}=\frac{1}{4\pi^{2}v}\varepsilon
^{\varepsilon_{1}^{\ast}\varepsilon_{2}^{\ast}q_{1}q_{2}}\mathcal{N}_{V_1,V_2}\ ,\ \ \ \varepsilon^{\varepsilon_{1}^{\ast}\varepsilon_{2}^{\ast}%
q_{1}q_{2}}\equiv\varepsilon^{\alpha\beta\mu\nu}\varepsilon(q_{1})_{\alpha
}^{\ast}\varepsilon(q_{2})_{\beta}^{\ast}q_{1\mu}q_{2\nu}\ ,
\end{equation}
for some mass-independent coefficients $\mathcal{N}_{V_1,V_2}$. Working out their values, to these amplitudes correspond the effective interactions%
\begin{align}
\mathcal{L}_{\mathrm{Linear}}^{\mathrm{eff}} &  =\frac{A^{0}}{16\pi^{2}%
v}\left(  g_{s}^{2}\mathcal{N}_{C}G_{\mu\nu}^{a}\tilde{G}^{a,\mu\nu}%
+e^{2}\mathcal{N}_{em}F_{\mu\nu}\tilde{F}^{\mu\nu}+\frac{2e^{2}}{c_{W}s_{W}%
}\left(  \mathcal{N}_{0}-s_{W}^{2}\mathcal{N}_{em}\right)  Z_{\mu\nu}\tilde
{F}^{\mu\nu}\right. \nonumber\\
&  \ \ \ \ \ \ \ \ \ \ \ \ \ \ \ \ \ \ \ \ \ \ \left.  +\frac{e^{2}}{c_{W}%
^{2}s_{W}^{2}}\left(  \mathcal{N}_{1}-2s_{W}^{2}\mathcal{N}_{0}+s_{W}%
^{4}\mathcal{N}_{em}\right)  Z_{\mu\nu}\tilde{Z}^{\mu\nu}+2\mathcal{N}%
_{2}g^{2}W_{\mu\nu}^{+}\tilde{W}^{-,\mu\nu}\right)  \ ,\label{LinINFTY}%
\end{align}
with the coefficients (to simplify the comparison, we adopt the standard convention $x\equiv1/\tan\beta$):
\begin{subequations}
\label{CoeffLin}%
\begin{align}
\mathcal{N}_{C} &  =\frac{1}{2}\left(  x+\frac{1}{x}\right)  \ ,\\
\mathcal{N}_{em} &  =N_{C}\left(  \frac{4x}{9}+\frac{1}{9x}\right)  +\frac
{1}{x}\ ,\\
\mathcal{N}_{0} &  =\frac{1}{4}\left(  N_{C}\left(  \frac{2x}{3}+\frac{1}%
{3x}\right)  +\frac{1}{x}\right)  \ ,\\
\mathcal{N}_{1} &  =\frac{1}{12}\left(  N_{C}\left(  x+\frac{1}{x}\right)
+\frac{1}{x}\right)  \ ,\\
\mathcal{N}_{2} &  =\frac{1}{12}\left(  N_{C}\left(  x+\frac{1}{x}\right)
+\frac{3}{2x}\right)  \ .
\end{align}
We stress that these effective couplings are directly derived from the well-established $A^{0}\rightarrow V_{1}V_{2}$ amplitudes computed in the standard THDM\footnote{These results apply to axion like particles but also to heavier pseudoscalars.}, as obtained in Ref.~\cite{Gunion:1991cw}.

Often, one interprets the non-decoupling of these amplitudes when $m_{u,d,e}\rightarrow\infty$ as an indirect manifestation of the underlying anomalies, but it is important to understand this statement correctly. The pseudoscalar triangles do decouple in that limit, $\mathcal{T}_{PVV}^{\alpha\beta}(m_{f})\rightarrow0$ and $\mathcal{T}_{PAA}^{\alpha\beta}(m_{f})\rightarrow0$ when $m_{f}\rightarrow\infty$, but this is compensated by the mass-dependent couplings of the $A^{0}$ to fermions. The fact that this compensation works, leaving a constant remainder in the $m_{f}\rightarrow \infty$, is guaranteed by the anomalous axial Ward identity of Eq.~(\ref{AxialWard}). Yet, there is no anomalous contribution of any kind to the $A^{0}\rightarrow V_{1}V_{2}$ amplitudes, and those are driven entirely by non-local and well-behaved fermion loops. The situation is quite similar for the loop-induced Higgs decay $h\rightarrow\gamma\gamma$ in the SM, whose non-decoupling behavior in the $m_{u,d,e}\rightarrow\infty$ limit is interpreted as a manifestation of the trace anomaly~\cite{Djouadi:2005gi}. Yet, there is of course no direct contribution of that anomaly to $h\rightarrow\gamma\gamma$, which can be safely computed perturbatively.

\subsection{Polar representation}

The Higgs boson fields are written in the polar representation\footnote{Note that one cannot start from the rotated basis Eq.~(\ref{HiggsBasis}) to exponentiate the Goldstone field $A^{0}$, because such a representation would not be canonical for $\Phi_{H}$ since it has no vacuum expectation value~\cite{Kamefuchi:1961sb}.} as (see Eq.~(\ref{PolarP})):
\end{subequations}
\begin{equation}
\Phi_{1}=\frac{1}{\sqrt{2}}\exp(i\eta_{1}/v_{1})\left( \begin{array}[c]{c} H_{1}^{+}\\ v_{1}+\operatorname{Re}H_{1} \end{array} \right)  \;,\;
\Phi_{2}=\frac{1}{\sqrt{2}}\exp(i\eta_{2}/v_{2})\left( \begin{array}[c]{c} H_{2}^{+}\\ v_{2}+\operatorname{Re}H_{2}\end{array} \right)  \;,
\end{equation}
where $v_{1}=v\sin\beta$, $v_{2}=v\cos\beta$. The $\eta_{1}$ and $\eta_{2}$ states are the Goldstone bosons associated to
the $U(1)_{1}$ and $U(1)_{2}$ symmetries. They are related to the physical
Goldstone bosons $a^{0}$ and $G^{0}$ by the same rotation as in
Eq.~(\ref{RotBeta}):%
\begin{equation}
\left(
\begin{array}[c]{c} G^{0}\\ a^{0} \end{array} \right)  =\left(
\begin{array}[c]{cc} 
\cos\beta & \sin\beta\\ 
-\sin\beta & \cos\beta
\end{array}
\right)  \left(
\begin{array}[c]{c} \eta_{2}\\ \eta_{1} \end{array} \right)  \ .
\end{equation}
Note that the $G^{0}$ components of $\Phi_{1}$ and $\Phi_{2}$ are the same since they have the same hypercharge, and the $a^{0}$ and $G^{0}$ states are orthogonal to each other. Now, the $G^{0}$ can be gauged away via a $U(1)_{Y}$ rotation, leaving%
\begin{equation}
\Phi_{1}=\frac{1}{\sqrt{2}}\exp(ia^{0}x/v)\left(
\begin{array}[c]{c} H_{1}^{+}\\ v_{1}+\operatorname{Re}H_{1} \end{array}
\right)  \;,\;\Phi_{2}=\frac{1}{\sqrt{2}}\exp(-ia^{0}/xv)\left(
\begin{array}[c]{c} H_{2}^{+}\\ v_{2}+\operatorname{Re}H_{2}\end{array} \right)  \;,
\end{equation}
where we introduce $x=v_{2}/v_{1}=1/\tan\beta$ to stick to common conventions. The remaining neutral Goldstone field, here denoted $a^{0}$ instead of $A^{0}$, is the axion. Under this form, the true $U(1)_{PQ}$ charges of the Higgs doublets are finally apparent, since under such a transformation,
$a^{0}\rightarrow a^{0}+v\theta$. Then, the Yukawa Lagrangian can also be made invariant under $U(1)_{PQ}$ provided the fermion charges are set as%
\begin{equation}
\Phi_{1}\rightarrow\exp(i\theta x)\Phi_{1}\ ,\ \ \Phi_{2}\rightarrow
\exp(-i\theta/x)\Phi_{2}\ ,\ \ \psi\rightarrow\exp(i\chi_{\psi}\theta
)\psi\ ,\label{PQtf}%
\end{equation}
with%
\begin{equation}
\chi_{q_{L}}=\alpha\ ,\ \chi_{u_{R}}=\alpha+x\ ,\ \chi_{d_{R}}=\alpha+\frac
{1}{x}\ ,\ \chi_{\ell_{L}}=\beta\ ,\ \chi_{e_{R}}=\beta+\frac{1}%
{x}\ ,\label{PQferm}%
\end{equation}
where $\alpha$ and $\beta$ are free parameters, corresponding to the conservation of baryon and lepton numbers $\mathcal{B}$ and $\mathcal{L}$.

Following the same steps as in Section~\ref{Toy}, we now perform a reparametrization of the fermion fields to force them to be invariant under $U(1)_{PQ}$. This is achieved by the field-dependent rotations $\psi\rightarrow\exp(i\chi_{\psi}a^{0}(x)/v)\psi$. The axion field disappears from the Yukawa Lagrangian, but reappears in derivative interactions from the fermion kinetic terms%
\begin{equation}
\delta\mathcal{L}_{\text{\textrm{Der}}}=-\frac{\partial_{\mu}a^{0}}{v}%
\sum_{\psi=q_{L},u_{R},d_{R},\ell_{L},e_{R}}\chi_{\psi}\bar{\psi}\gamma^{\mu
}\psi=-\frac{\partial_{\mu}a^{0}}{2v}\sum_{f=u,d,e,\nu}\left(  \chi_{V}%
^{f}~\bar{\psi}_{f}\gamma^{\mu}\psi_{f}+\chi_{A}^{f}~\bar{\psi}_{f}\gamma
^{\mu}\gamma_{5}\psi_{f}\frac{{}}{{}}\right)  \ ,\label{LDer}%
\end{equation}
with
\begin{equation}%
\begin{array}
[c]{llll}%
\chi_{V}^{u}=2\alpha+x\ , & \chi_{V}^{d}=2\alpha+\dfrac{1}{x}\ , & \chi
_{V}^{e}=2\beta+\dfrac{1}{x}\ , & \chi_{V}^{\nu}=\beta\ ,\\
\chi_{A}^{u}=x\ , & \chi_{A}^{d}=\dfrac{1}{x}\ , & \chi_{A}^{e}=\dfrac{1}%
{x}\ , & \chi_{A}^{\nu}=-\beta\ .
\end{array}
\label{chiAV}%
\end{equation}
The vector current couplings are $\alpha $ and $\beta $-dependent, since those parameters reflect the conservation of $\mathcal{B}$ and $\mathcal{L}$, which are vectorial~\cite{Georgi:1986df}. The $\beta $ dependence of $\chi_{A}^{\nu }$ is related to the peculiar nature of neutrinos, which are kept massless. As we will see in Section~\ref{neutrino}, it disappears if right handed neutrinos $\nu_{R}$ together with the $\mathcal{L}$-invariant Yukawa interaction $\bar{\nu}_{R}Y_{\nu }\ell _{L}\Phi _{1}$ are added.

In addition, given the $U(1)_{PQ}$ charges in Eq.~(\ref{PQtf}), the fermion reparametrization is anomalous and its Jacobian has to be included%
\begin{equation}
\delta\mathcal{L}_{\text{\textrm{Jac}}}=\frac{a^{0}}{16\pi^{2}v}\left(
g_{s}^{2}\mathcal{N}_{C}G_{\mu\nu}^{a}\tilde{G}^{a,\mu\nu}+g^{2}%
\mathcal{N}_{L}W_{\mu\nu}^{i}\tilde{W}^{i,\mu\nu}+g^{\prime2}\mathcal{N}%
_{Y}B_{\mu\nu}\tilde{B}^{\mu\nu}\frac{{}}{{}}\right)  \;.\label{LJac}%
\end{equation}
Given the SM fermion quantum numbers, the coefficients are%
\begin{align}
\mathcal{N}_{C} &  =\sum_{\psi=q_{L},u_{R}^{\dagger},d_{R}^{\dagger}}%
d_{L}^{\psi}C_{C}^{\psi}\chi_{\psi}=\frac{1}{2}(-2\chi_{q_{L}}+\chi_{u_{R}%
}+\chi_{d_{R}})=\frac{1}{2}\left(  x+\frac{1}{x}\right)  \ ,\\
\mathcal{N}_{L} &  =\sum_{\psi=q_{L},\ell_{L}}d_{C}^{\psi}C_{L}^{\psi}%
\chi_{\psi}=\frac{1}{2}(-N_{C}\chi_{q_{L}}-\chi_{\ell_{L}})=-\frac{1}{2}%
(N_{C}\alpha+\beta)\ ,\\
\mathcal{N}_{Y} &  =\sum_{\psi=q_{L},u_{R}^{\dagger},d_{R}^{\dagger},\ell
_{L},e_{R}^{\dagger}}d_{L}^{\psi}d_{C}^{\psi}C_{Y}^{\psi}\chi_{\psi}=\frac
{1}{2}\left(  N_{C}\alpha+\beta\right)  +N_{C}\left(  \frac{4}{9}x+\frac
{1}{9x}\right)  +\frac{1}{x}\ ,
\end{align}
where $d_{C,L}^{\psi}$, $C_{C,L}^{\psi}$ are the $SU(3)_{C}$ and $SU(2)_{L}$ dimensions and quadratic Casimir invariant of the representation carried by the field $\psi$, respectively, and by extension, $C_{Y}^{\psi}=Y(\psi)^{2}/4$ with $Y(q_{L},u_{R},d_{R},\ell_{L},e_{R})=(1/3,4/3,-2/3,-1,-2)$. These expressions correspond to Eq.~(\ref{AnoLLL}), choosing $a=-b=1$ to reflect the absence of anomalies for the SM gauge symmetries (so no Bose symmetry here).

The parameters $\alpha$ and $\beta$ enter in a combination reflecting the anomaly in the $\mathcal{B}+\mathcal{L}$ current. Interestingly, neither the QED nor the QCD coefficients depend on them, but those to electroweak gauge bosons do. To see this, it sufficies to plug $W_{\mu}^{3}=c_{W}Z_{\mu}+s_{W}A_{\mu}$, $B_{\mu}=-s_{W}Z_{\mu}+c_{W}A_{\mu}$ and $g^{\prime}c_{W}=e=gs_{W}$ in Eq.~(\ref{LJac}):
\begin{align}
\delta\mathcal{L}_{\text{\textrm{Jac}}} &  =\frac{a^{0}}{16\pi^{2}v}\left(
g_{s}^{2}\mathcal{N}_{C}G_{\mu\nu}^{a}\tilde{G}^{a,\mu\nu}+e^{2}%
\mathcal{N}_{em}F_{\mu\nu}\tilde{F}^{\mu\nu}+\frac{2e^{2}}{c_{W}s_{W}}\left(
\mathcal{N}_{L}-s_{W}^{2}\mathcal{N}_{em}\right)  Z_{\mu\nu}\tilde{F}^{\mu\nu
}\right. \nonumber\\
&  \ \ \ \ \ \ \ \ \ \ \ \ \ \ \ \ \ \left.  +\frac{e^{2}}{c_{W}%
^{2}s_{W}^{2}}\left(  (1-2s_{W}^{2})\mathcal{N}_{L}+s_{W}^{4}\mathcal{N}%
_{em}\right)  Z_{\mu\nu}\tilde{Z}^{\mu\nu}+2\mathcal{N}_{L}g^{2}W_{\mu\nu}%
^{+}\tilde{W}^{-,\mu\nu}\right)  \ ,\label{DLJac}%
\end{align}
and the free parameters cancel out in
\begin{equation}
\mathcal{N}_{em}\equiv\mathcal{N}_{L}+\mathcal{N}_{Y}=N_{C}\left(  \frac{4}%
{9}x+\frac{1}{9x}\right)  +\frac{1}{x}\ .
\end{equation}

Concerning the strong $CP$ puzzle, the polar representation makes the shift symmetry explicit. Under a $U(1)_{PQ}$ transformation of parameter $\theta$, all the fields are invariant but for $a^{0}\rightarrow a^{0}+v\theta$. So, if there is a $\theta_{QCD}G_{\mu\nu}^{a}\tilde{G}^{a,\mu\nu}$ term in the gauge sector, it can be eliminated by an adequate shift of the $a^{0}$ field. As said in the introduction, what happens in reality is that one is forced by non-perturbative QCD effects to shift the $a^{0}$ field by precisely that amount~\cite{PQ}. Once the $a^{0}$ shift is set by QCD, the electroweak $\theta_{L}W_{\mu\nu}^{i}\tilde{W}^{i,\mu\nu}$ can still be removed by an appropriate choice of $N_{C}\alpha+\beta$. All that remains then is the abelian $\theta_{Y}B_{\mu\nu}\tilde{B}^{\mu\nu}$, which is an inoccuous total derivative. We thus recover the well-known fact that the electroweak $\theta$ term can be rotated away thanks to the $\mathcal{B}$ and $\mathcal{L}$ invariance of the model.

\subsection{Matching the polar and linear representations}

The physics must not depend on the representation chosen for the scalar fields or on the parametrization of the fermion fields. So, the one-loop $A^{0}\rightarrow V_{1}V_{2}$ amplitudes of Eq.~(\ref{LinAmpli}) must match the $a^{0}\rightarrow V_{1}V_{2}$ amplitudes computed using the local terms $\delta\mathcal{L}_{\text{\textrm{Jac}}}$ together with the triangle graphs arising from the derivative interactions in $\delta \mathcal{L}_{\text{\textrm{Der}}}$ (these two contributions are of the same order in the coupling constants). At this stage, it is immediately clear that the triangle graphs will play a crucial role. Indeed, taken alone, the contributions from the anomalous local terms of $\delta\mathcal{L}_{\text{\textrm{Jac}}}$ match those found in the linear case in the $m_{u,d,e}\rightarrow\infty$ limit, Eq.~(\ref{LinINFTY}), only for the $\gamma\gamma$ and $gg$ final states. For the others, there is no dependence on $\alpha$ and $\beta$ in Eq.~(\ref{LinINFTY}), but there is one through $\mathcal{N}_{L}$ in $\delta\mathcal{L}_{\text{\textrm{Jac}}}$. Even worse, no choice of these parameters could make the two results compatible because the relative strengths of the $\gamma\gamma$, $\gamma Z$, $ZZ$, and $W^{+}W^{-}$ decays are irremediably different. As we will now detail, it is only once the anomalous contributions from $\delta\mathcal{L}_{\text{\textrm{Jac}}}$ precisely cancel out with those hidden in the triangle graphs that the $a^{0}\rightarrow V_{1}V_{2}$ amplitudes match the $A^{0}\rightarrow V_{1}V_{2}$ amplitudes.

Throughout this section, the momentum flow is defined as $a^{0}(q_{1}+q_{2})\rightarrow V_{1}(q_{1},\alpha)V_{2}(q_{2},\beta)$, with $V_{1,2}=g,\gamma,Z,W$, and neither the final gauge bosons nor the initial axion are necessarily on-shell. The final amplitudes thus depend on $q_{1}^{2}$, $q_{2}^{2}$, and $(q_{1}+q_{2})^{2}$, and are to be compared to those obtained in the linear case in Eq.~(\ref{LinAmpli}).

\subsubsection{The $a^{0}\rightarrow\gamma\gamma$ and $a^{0}\rightarrow gg$ decays}

\begin{figure}[t]
\centering\includegraphics[width=0.95\textwidth]{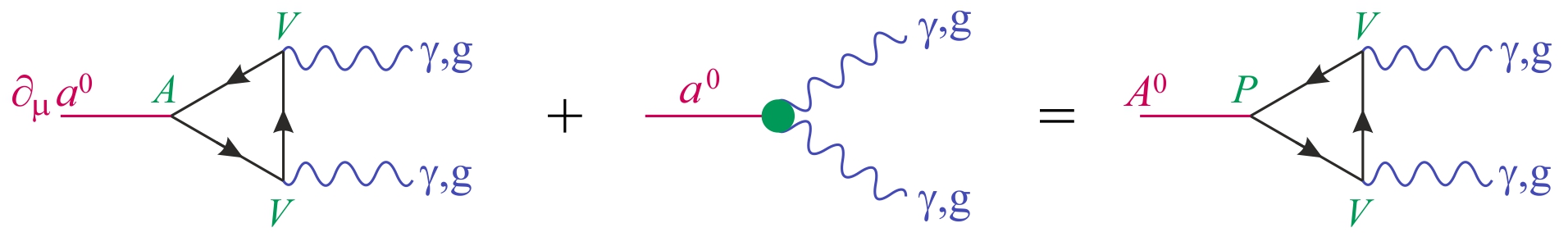}
\caption{Representation of the contributions to $a^{0}\rightarrow\gamma\gamma,gg$ in the polar representation, and their matching with the $A^{0}\rightarrow\gamma\gamma,gg$ amplitude of the linearly realized theory. The notation $P,V,A$ denotes pseudoscalar, vector, and axial vertices, that is, $\gamma_{5}$, $\gamma^{\mu}$, and $\gamma^{\mu}\gamma_{5}$ Dirac structures. The green disk depicts the local anomalous vertex derived from Eq.~(\ref{LJac}). All the SM fermions but the neutrinos circulate in the $\gamma\gamma$ loops, while only quarks occur for the $gg$ loops.}%
\label{Fig3a}%
\end{figure}

For the decays into photons or gluons, only the $AVV$ triangle contributes since $SU(3)$ and $U(1)_{em}$ are vector symmetries. Then, the situation is exactly the same as in the toy model of Section~\ref{Toy}. Setting $a=-b=1$ in Eq.~(\ref{AnoAVVm}), we find
\begin{align}
\mathcal{M}(a^{0}\overset{}{\rightarrow}\gamma\gamma)_{\text{\textrm{Der}}%
}^{AVV}  & =-\frac{e^{2}}{2v}\sum_{f=u,d,e}d_{C}^{f}\times\chi_{A}^{f}%
Q_{f}^{2}\times i(q_{1}+q_{2})_{\gamma}\mathcal{T}_{AVV}^{\gamma\alpha\beta
}(f)\nonumber\\
& =-\mathcal{M}(a^{0}\overset{}{\rightarrow}\gamma\gamma)_{\text{\textrm{Jac}%
}}+\mathcal{M}(A^{0}\overset{}{\rightarrow}\gamma\gamma)_{\mathrm{Linear}}\ ,
\end{align}
and
\begin{align}
\mathcal{M}(a^{0}\overset{}{\rightarrow}gg)_{\text{\textrm{Der}}}^{AVV}  &
=-\frac{g_{s}^{2}}{2v}\sum_{f=u,d}\chi_{A}^{f}C_{C}^{f}\times i(q_{1}%
+q_{2})_{\gamma}\mathcal{T}_{AVV}^{\gamma\alpha\beta}(f)\nonumber\\
& =-\mathcal{M}(a^{0}\overset{}{\rightarrow}gg)_{\text{\textrm{Jac}}%
}+\mathcal{M}(A^{0}\overset{}{\rightarrow}gg)_{\mathrm{Linear}}\ ,
\end{align}
where $d_{C}^{u,d}=N_{C}$, $d_{C}^{e,\nu}=1$, $Q^{f}$ is the electric charge
of $f$, and $C_{C}^{u,d}=1/2$, $C_{C}^{e,\nu}=0$. The amplitudes in the polar
representation thus trivially match that in the linear case, see
Fig.~\ref{Fig3a},
\begin{equation}
\mathcal{M}(a^{0}\overset{}{\rightarrow}\gamma\gamma,gg)_{_{\mathrm{Polar}}%
}=\mathcal{M}(a^{0}\overset{}{\rightarrow}\gamma\gamma,gg)_{\text{\textrm{Der}%
}}^{AVV}+\mathcal{M}(a^{0}\overset{}{\rightarrow}\gamma\gamma
,gg)_{\text{\textrm{Jac}}}=\mathcal{M}(A^{0}\overset{}{\rightarrow}%
\gamma\gamma,gg)_{\mathrm{Linear}}\ .
\end{equation}
In other words, the anomalous contact interactions do cancel out systematically with the anomalous part of the triangle graphs.%

As for the toy model, the polar representation is thus interesting only to make the shift symmetry manifest, and because the contact $a^{0}\gamma\gamma$ and $a^{0}gg$ interactions read off $\delta\mathcal{L}_{\text{\textrm{Jac}}}$ are reliable book-keeping of the effects of heavy fermions. Specifically, $\mathcal{M}(a^{0}\rightarrow\gamma\gamma,gg)_{\text{\textrm{Der}}}^{AVV}\overset{m\rightarrow\infty}{=}0$ implies that $\mathcal{M}(A^{0}\overset{}{\rightarrow}\gamma\gamma,gg)_{\mathrm{Linear}}\overset{m\rightarrow\infty}{=}\mathcal{M}(a^{0}\overset{}{\rightarrow}\gamma\gamma,gg)_{\text{\textrm{Jac}}}$. Finally, remark that the cancellation of the local anomalous terms ensures $\mathcal{M}(a^{0}\overset{}{\rightarrow}\gamma\gamma,gg)_{_{\mathrm{Polar}}}=0$ in the $m_{u,d,e}\rightarrow0$ limit. So, though interpreting the axion coupling to photons or gluons as induced by the anomaly is incorrect, this misidentification does not lead to serious consequences for those final states.\ For heavy fermions, the coupling to gluons is tuned by $\mathcal{N}_{C}$, and that to photons by $\mathcal{N}_{em}$, and their ratio, when restricted to quarks, give back the usual $\mathcal{N}_{em}^{q}/\mathcal{N}_{C}=8/3$. However, as we will see in the next subsection, interpreting the axion coupling involving at least one electroweak gauge boson as induced by the anomaly is not only wrong in principle but also leads to incorrect couplings.

\subsubsection{The $a^{0}\rightarrow\gamma Z$ decay}

\begin{figure}[t]
\centering\includegraphics[width=0.95\textwidth]{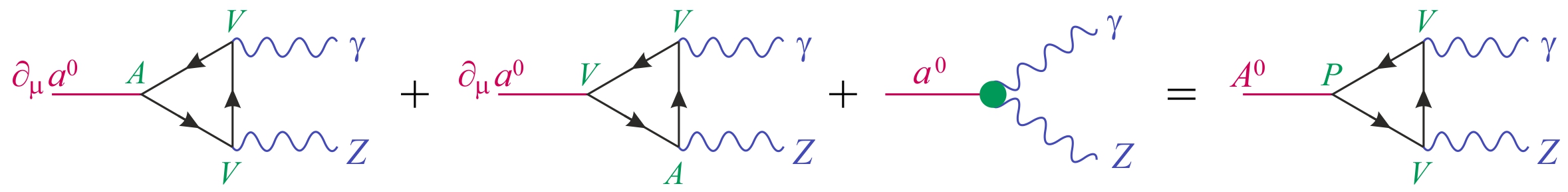}
\caption{Contributions to the axion decay into $\gamma Z$ in the polar (left) and linear (right) representations, using the same notations as in Fig.~\ref{Fig3a}. All the SM fermions but the neutrinos circulate in the loops.}%
\label{Fig3b}%
\end{figure}

For electroweak gauge bosons in the final state, the situation is less simple. Consider first the $\gamma Z$ final state. The derivative interactions induce again only the $AVV$ triangle graphs since the photon coupling is vectorial. However, this time we have two possible contributions, depending on which current is carrying the anomaly, see Fig.~\ref{Fig3b}. First, there are the usual $A(\partial_{\mu}a^{0})-V(\gamma)-V(Z)$ triangles, for which the anomaly is in the axial current. Using Eq.~(\ref{AnoAVVm}) with $a=-b=1$, we find
\begin{align}
\mathcal{M}(a^{0}\overset{}{\rightarrow}\gamma Z)_{\text{\textrm{Der}}}^{AVV}
& =-\frac{1}{2v}\sum_{f=u,d,e}d_{C}^{f}\times\chi_{A}^{f}g_{V,\gamma}%
^{f}g_{V,Z}^{f}\times i(q_{1}+q_{2})_{\gamma}\mathcal{T}_{AVV}^{\gamma
\alpha\beta}(f)\nonumber\\
& =-\frac{1}{4\pi^{2}v}\frac{e^{2}}{c_{W}s_{W}}\left(  \mathcal{N}_{0}%
-s_{W}^{2}\mathcal{N}_{em}\right)  \varepsilon^{\varepsilon_{1}^{\ast
}\varepsilon_{2}^{\ast}q_{1}q_{2}}+\mathcal{M}(A^{0}\overset{}{\rightarrow
}\gamma Z)_{\mathrm{Linear}}\ ,
\end{align}
where $\varepsilon^{\varepsilon_{1}^{\ast}\varepsilon_{2}^{\ast}q_{1}q_{2}}\equiv\varepsilon^{\alpha\beta\mu\nu}\varepsilon(q_{1})_{\alpha}^{\ast}\varepsilon(q_{2})_{\beta}^{\ast}q_{1\mu}q_{2\nu}$. But contrary to the $\gamma\gamma$ and $gg$ final state, there are now new contributions from the $V(\partial_{\mu}a^{0})-V(\gamma)-A(Z)$ triangles, with the axion vector couplings of Eq.~(\ref{LDer}). Using again Eq.~(\ref{AnoAVVm}) but this time with $a=b=1$ to preserve $SU(2)_{L}\otimes U(1)_{Y}$, and noting that
$-i(q_{1})_{\gamma}\mathcal{T}_{AVV}^{\alpha\gamma\beta}=-i(q_{1}+q_{2})_{\gamma}\mathcal{T}_{VAV}^{\gamma\alpha\beta}$,
\begin{align}
\mathcal{M}(a^{0}\overset{}{\rightarrow}\gamma Z)_{\text{\textrm{Der}}}^{VAV}
& =-\frac{1}{2v}\sum_{f=u,d,e}d_{C}^{f}\times\chi_{V}^{f}g_{V,\gamma}%
^{f}g_{A,Z}^{f}\times (-i)(q_{1}+q_{2})_{\gamma}\mathcal{T}_{VAV}^{\gamma
\alpha\beta}\nonumber\\
& =\frac{1}{4\pi^{2}v}\frac{e^{2}}{c_{W}s_{W}}\left(  \mathcal{N}%
_{0}-\mathcal{N}_{L}\right)  \varepsilon^{\varepsilon_{1}^{\ast}%
\varepsilon_{2}^{\ast}q_{1}q_{2}}\ .
\end{align}
As explained in Section~\ref{Ano}, this contribution is free of any mass-dependent term because the naive vector current conservation must hold, up to the anomaly.%

These two contributions from the derivative interactions combine with the Jacobian term from Eq.~(\ref{DLJac}),%
\begin{equation}
\mathcal{M}(a^{0}\overset{}{\rightarrow}\gamma Z)_{\text{\textrm{Jac}}}%
=\frac{1}{4\pi^{2}v}\frac{e^{2}}{c_{W}s_{W}}\left(  \mathcal{N}_{L}-s_{W}%
^{2}\mathcal{N}_{em}\right)  \varepsilon^{\varepsilon_{1}^{\ast}%
\varepsilon_{2}^{\ast}q_{1}q_{2}}\ ,
\end{equation}
to give the total decay amplitude in the polar representation:%
\begin{align}
\mathcal{M}(a^{0}\overset{}{\rightarrow}\gamma Z)_{\text{\textrm{Polar}}}  &
=\mathcal{M}(a^{0}\overset{}{\rightarrow}\gamma Z)_{\text{\textrm{Der}}}%
^{AVV}+\mathcal{M}(a^{0}\overset{}{\rightarrow}\gamma Z)_{\text{\textrm{Der}}%
}^{VAV}+\mathcal{M}(a^{0}\overset{}{\rightarrow}\gamma Z)_{\text{\textrm{Jac}%
}}\nonumber\\
& =\mathcal{M}(A^{0}\overset{}{\rightarrow}\gamma Z)_{\mathrm{Linear}}\ ,
\end{align}
as it should. 
We can now understand why $\mathcal{M}(a^{0}\overset{}{\rightarrow}\gamma Z)_{\text{\textrm{Jac}}}$ does not match $\mathcal{M}(A^{0}\overset{}{\rightarrow}\gamma Z)_{\mathrm{Linear}}$ in the $m_{u,d,e}\rightarrow\infty$ limit. Indeed, while
$\mathcal{M}(a^{0}\rightarrow\gamma Z)_{\text{\textrm{Der}}}^{AVV}$ vanishes in the $m_{u,d,e}\rightarrow\infty$ limit, $\mathcal{M}(a^{0}\overset{}{\rightarrow}\gamma Z)_{\text{\textrm{Der}}}^{VAV}$ obviously does not since it is independent of $m_{u,d,e}$. In that limit, we should write
\begin{equation}
\left.  \mathcal{M}(A^{0}\overset{}{\rightarrow}\gamma Z)_{\mathrm{Linear}%
}\right\vert _{m_{u,d,e}\rightarrow\infty}=\mathcal{M}(a^{0}\overset
{}{\rightarrow}\gamma Z)_{\text{\textrm{Der}}}^{VAV}+\mathcal{M}(a^{0}%
\overset{}{\rightarrow}\gamma Z)_{\text{\textrm{Jac}}}\ ,
\end{equation}
in which the parameters $\alpha$ and $\beta$ cancel out. So, for a chiral gauge theory, the local terms coming from $\delta\mathcal{L}_{\text{\textrm{Jac}}}$ are no longer reliable book-keeping of the effect of heavy fermions, because part of the anomaly is hidden in the $VAV$ triangle.

Actually, the reason why the $VAV$ triangle plays a role can be understood directly from the fermion reparametrization, given the charges in Eq.~(\ref{PQtf}). When $\alpha$ and/or $\beta$ are different from zero, the fermionic current associated to $U(1)_{PQ}$ has a component aligned with baryon and/or lepton number, respectively. The fermionic reparametrization thus generates the anomalous $\mathcal{B}+\mathcal{L}$ interactions: those correspond to the $N_{C}\alpha+\beta$ terms of $\mathcal{N}_{L}$ and $\mathcal{N}_{Y}$ in Eq.~(\ref{LJac}). But as for the toy model, these interactions are spurious and must cancel with the anomalies present in the triangle graphs induced by the derivative interactions $\delta\mathcal{L}_{\text{\textrm{Der}}}$. This must necessarily come from a breakdown of the vector Ward identity for the $VAV$ triangle since $\mathcal{B}$ and $\mathcal{L}$ are purely vectorial symmetries~\cite{Dreiner:2008tw}.

\subsubsection{The $a^{0}\rightarrow ZZ$ and $a^{0}\rightarrow W^{+}W^{-}$ decays}

\begin{figure}[t]
\centering\includegraphics[width=0.95\textwidth]{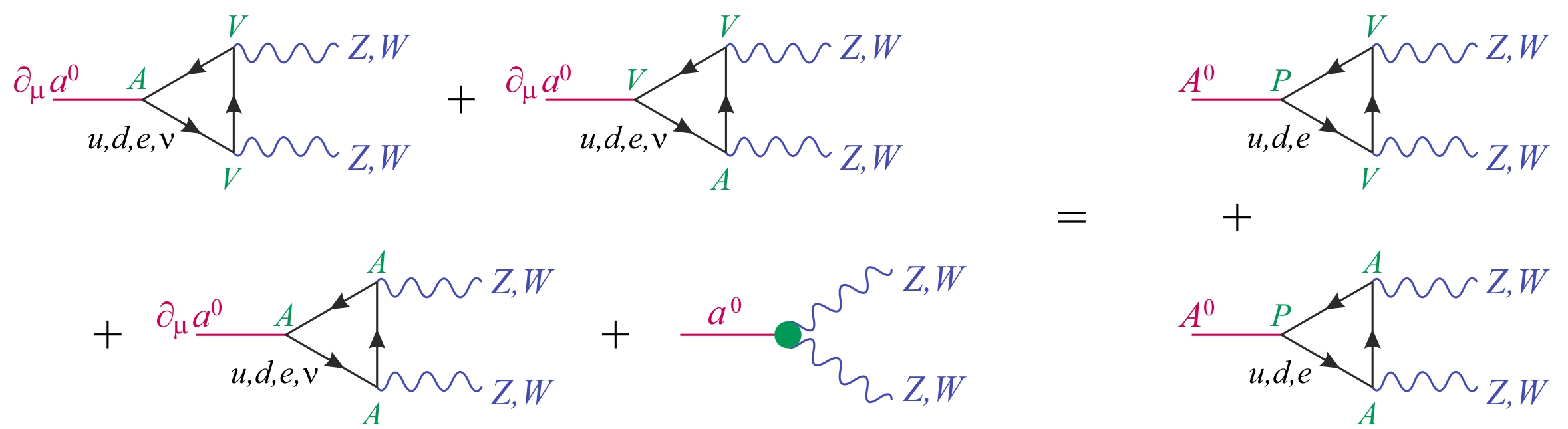}
\caption{Contributions to the axion decay into $ZZ$ and $W^{+}W^{-}$ in the polar (left) and linear (right) representations, using the same notations as in Fig.~\ref{Fig3a}. For $ZZ$, all the SM fermions circulate in the loops in the non-linear theory, but the neutrino is absent in the linear one since we take it as massless. For $W^{+}W^{-}$, it is understood that in between the two $W$s, the $SU(2)_{L}$ partner of the indicated fermion propagates.}%
\label{Fig3c}%
\end{figure}

Turning first to the $a^{0}\rightarrow ZZ$ amplitude, the derivative interactions induce new types of diagrams: the $AAA$ triangles and graphs with neutrinos circulating in the loop, see Fig.~\ref{Fig3c}.%

Proceeding with the calculation, the three contributions from the derivative terms are%
\begin{align}
\mathcal{M}(a^{0}\overset{}{\rightarrow}ZZ)_{\text{\textrm{Der}}}^{AVV} &
=-\frac{1}{2v}\sum_{f=u,d,e,\nu}d_{C}^{f}\times\chi_{A}^{f}g_{V,Z}^{f}%
g_{V,Z}^{f}\times i(q_{1}+q_{2})_{\gamma}\mathcal{T}_{AVV}^{\gamma\alpha\beta
}(f)\nonumber\\
&  =\frac{-1}{4\pi^{2}v}\frac{e^{2}}{c_{W}^{2}s_{W}^{2}}\left(  \frac{3}%
{4}\mathcal{N}_{1}-2s_{W}^{2}\mathcal{N}_{0}+s_{W}^{4}\mathcal{N}_{em}%
-\frac{\beta}{16}\right)  \varepsilon^{\varepsilon_{1}^{\ast}\varepsilon
_{2}^{\ast}q_{1}q_{2}}+\mathcal{M}(A^{0}\overset{}{\rightarrow}%
ZZ)_{\mathrm{Linear}}^{PVV},
\end{align}
where $\mathcal{M}(A^{0}\overset{}{\rightarrow}Z Z)_{\mathrm{Linear}}^{PVV}$ is the part of Eq.~(\ref{LinAmpli}) proportional to $\mathcal{T}_{PVV}^{\alpha\beta}$. The neutrino contribution is absent from that term since the loop function in Eq.~(\ref{TriPVVPAA}) vanishes for massless fermions. The next piece comes from the divergence of the axionic vector current,
\begin{align}
\mathcal{M}(a^{0}\overset{}{\rightarrow}ZZ)_{\text{\textrm{Der}}}^{VAV}  &
=-\frac{1}{2v}\sum_{f=u,d,e,\nu}d_{C}^{f}\times\chi_{V}^{f}(g_{V,Z}^{f}%
g_{A,Z}^{f}+g_{A,Z}^{f}g_{V,Z}^{f})\times (- i)(q_{1}+q_{2})_{\gamma}%
\mathcal{T}_{VAV}^{\gamma\alpha\beta}\nonumber\\
& =\frac{1}{4\pi^{2}v}\frac{e^{2}}{c_{W}^{2}s_{W}^{2}}\left(  \frac{3}%
{2}\mathcal{N}_{1}-2s_{W}^{2}\mathcal{N}_{0}-(1-2s_{W}^{2})\mathcal{N}%
_{L}-\frac{\beta}{8}\right)  \varepsilon^{\varepsilon_{1}^{\ast}%
\varepsilon_{2}^{\ast}q_{1}q_{2}}\ .
\end{align}
There is no mass-dependent terms here because of the vector Ward identity. Finally, the $AAA$ triangle gives, using Eq.~(\ref{AnoAAAm}) with $a=-b=1$%
\begin{align}
\mathcal{M}(a^{0}\overset{}{\rightarrow}ZZ)_{\text{\textrm{Der}}}^{AAA}  &
=-\frac{1}{2v}\sum_{f=u,d,e,\nu}d_{C}^{f}\times\chi_{A}^{f}g_{A,Z}^{f}%
g_{A,Z}^{f}\times i(q_{1}+q_{2})_{\gamma}\mathcal{T}_{AAA}^{\gamma\alpha\beta
}\nonumber\\
& =\frac{-1}{4\pi^{2}v}\frac{e^{2}}{c_{W}^{2}s_{W}^{2}}\left(  \frac{3}%
{4}\mathcal{N}_{1}-\frac{\beta}{16}\right)  \varepsilon^{\varepsilon_{1}%
^{\ast}\varepsilon_{2}^{\ast}q_{1}q_{2}}+\mathcal{M}(A^{0}\overset
{}{\rightarrow}ZZ)_{\mathrm{Linear}}^{PAA}\ ,
\end{align}
where $\mathcal{M}(A^{0}\overset{}{\rightarrow} Z Z)_{\mathrm{Linear}}^{PAA}$ is the part of Eq.~(\ref{LinAmpli}) proportional to $\mathcal{T}_{PAA}^{\alpha\beta}$. Again, the neutrino contribution to that part vanishes. Combining the three contributions of the derivative interactions with that of the Jacobian%
\begin{equation}
\mathcal{M}(a^{0}\rightarrow Z Z)_{\text{\textrm{Jac}}}=\frac{1}{4\pi
^{2}v}\frac{e^{2}}{c_{W}^{2}s_{W}^{2}}\left(  (1-2s_{W}^{2})\mathcal{N}%
_{L}+s_{W}^{4}\mathcal{N}_{em}\right)  \varepsilon^{\varepsilon_{1}^{\ast
}\varepsilon_{2}^{\ast}q_{1}q_{2}}\ ,
\end{equation}
all the local anomalous contributions cancel exactly, leaving:%
\begin{align}
\mathcal{M}(a^{0}\overset{}{\rightarrow}ZZ)_{\text{\textrm{Polar}}}  &
=\mathcal{M}(a^{0}\overset{}{\rightarrow}ZZ)_{\text{\textrm{Der}}}%
^{AVV}+\mathcal{M}(a^{0}\overset{}{\rightarrow}ZZ)_{\text{\textrm{Der}}}%
^{VAV}\nonumber\\
& +\mathcal{M}(a^{0}\overset{}{\rightarrow}ZZ)_{\text{\textrm{Der}}}%
^{AAA}+\mathcal{M}(a^{0}\overset{}{\rightarrow}Z Z)_{\text{\textrm{Jac}}%
}\overset{}{=}\mathcal{M}(A^{0}\overset{}{\rightarrow}ZZ)_{\mathrm{Linear}}\ .
\end{align}

Though the anomalous terms again sum up to zero, they are now completely entangled in the three types of triangle graphs, which all depend on the free parameters $\alpha$ and $\beta$. As a result, none of the derivative amplitudes vanish in the $m_{u,d,e}\rightarrow\infty$ limit. Specifically, because neutrinos remain massless,
\begin{equation}
\left.  \mathcal{M}(a^{0}\rightarrow ZZ)_{\text{\textrm{Der}}}^{AVV}%
\right\vert _{m_{u,d,e}\rightarrow\infty}=-\frac{1}{4\pi^{2}v}\frac{e^{2}%
}{c_{W}^{2}s_{W}^{2}}\left(  -\frac{\beta}{16}\right)  \varepsilon
^{\varepsilon_{1}^{\ast}\varepsilon_{2}^{\ast}q_{1}q_{2}}\ ,
\end{equation}
where $\beta$ comes from the neutrino coupling $\chi_{A}^{\nu}$ in Eq.~(\ref{chiAV}). Also,
\begin{equation}
\left.  \mathcal{M}(a^{0}\rightarrow ZZ)_{\text{\textrm{Der}}}^{AAA}%
\right\vert _{m_{u,d,e}\rightarrow\infty}=-\frac{1}{4\pi^{2}v}\frac{e^{2}%
}{c_{W}^{2}s_{W}^{2}}\left(  \frac{1}{2}\mathcal{N}_{1}-\frac{\beta}%
{16}\right)  \varepsilon^{\varepsilon_{1}^{\ast}\varepsilon_{2}^{\ast}%
q_{1}q_{2}}\ ,
\end{equation}
because in addition to massless neutrinos, the mass term respects Bose symmetry but not the anomaly which resides only in the axion axial current (there is a relative factor of 1/3, see Eq.~(\ref{AnoAAAinf}) with $a=-b=1$). Finally, $\mathcal{M}(a^{0}\rightarrow ZZ)_{\text{\textrm{Der}}}^{VAV}$ does not vanish since it is mass-independent. Because of all this, $\mathcal{M}(a^{0}\overset{}{\rightarrow}ZZ)_{\text{\textrm{Jac}}}$ is clearly not identical to $\mathcal{M}(A^{0}\overset{}{\rightarrow}ZZ)_{\mathrm{Linear}}$ in the $m_{u,d,e}\rightarrow\infty$ limit, and cannot be used as an estimate of the electroweak couplings of the axion.

The $a^{0}\rightarrow W^{+}W^{-}$ amplitude shares many similarities with that for $a^{0}\rightarrow ZZ$, so we will be very brief. Following the same steps, the various contributions to the decay in the polar representation are%
\begin{align}
\mathcal{M}(a^{0}\overset{}{\rightarrow}W^{+}W^{-})_{\text{\textrm{Der}}%
}^{AVV}  & =-\frac{g^{2}}{4\pi^{2}v}\left(  \frac{3}{4}\mathcal{N}_{1}%
-\frac{\beta}{16}\right)  \varepsilon^{\varepsilon_{1}^{\ast}\varepsilon
_{2}^{\ast}q_{1}q_{2}}+\mathcal{M}(A^{0}\overset{}{\rightarrow}W^{+}%
W^{-})_{\mathrm{Linear}}^{PVV}\ ,\\
\mathcal{M}(a^{0}\overset{}{\rightarrow}W^{+}W^{-})_{\text{\textrm{Der}}%
}^{VAV}  & =+\frac{g^{2}}{4\pi^{2}v}\left(  \frac{3}{2}\mathcal{N}%
_{1}-\mathcal{N}_{L}-\frac{\beta}{8}\right)  \varepsilon^{\varepsilon
_{1}^{\ast}\varepsilon_{2}^{\ast}q_{1}q_{2}}\ ,\\
\mathcal{M}(a^{0}\overset{}{\rightarrow}W^{+}W^{-})_{\text{\textrm{Der}}%
}^{AAA}  & =-\frac{g^{2}}{4\pi^{2}v}\left(  \frac{3}{4}\mathcal{N}_{1}%
-\frac{\beta}{16}\right)  \varepsilon^{\varepsilon_{1}^{\ast}\varepsilon
_{2}^{\ast}q_{1}q_{2}}+\mathcal{M}(A^{0}\overset{}{\rightarrow}W^{+}%
W^{-})_{\mathrm{Linear}}^{PAA}\ ,\\
\mathcal{M}(a^{0}\overset{}{\rightarrow}W^{+}W^{-})_{\text{\textrm{Jac}}}  &
=+\frac{g^{2}}{4\pi^{2}v}\left(  \mathcal{N}_{L}\frac{{}}{{}}\right)
\varepsilon^{\varepsilon_{1}^{\ast}\varepsilon_{2}^{\ast}q_{1}q_{2}}\ .
\end{align}
Note well that it is $\mathcal{N}_{1}$ that occurs in these amplitudes, and not the $\mathcal{N}_{2}$ coefficients to which $\mathcal{M}(A^{0}\overset{}{\rightarrow}W^{+}W^{-})_{\mathrm{Linear}}$ tends to in the $m_{u,d,e}\rightarrow\infty$ limit. Still, the sum of these four contributions reproduces again the linear result, $\mathcal{M}(a^{0}\rightarrow W^{+}W^{-})_{\text{\textrm{Polar}}}=\mathcal{M}(A^{0}\rightarrow W^{+}W^{-})_{\mathrm{Linear}}$. As none of the triangle amplitudes vanish in the $m_{u,d,e}\rightarrow\infty$ limit, $\mathcal{M}(a^{0}\rightarrow W^{+}W^{-})_{\text{\textrm{Jac}}}$ is clearly not identical to $\mathcal{M}(A^{0}\rightarrow W^{+}W^{-})_{\mathrm{Linear}}$ in the $m_{u,d,e}\rightarrow\infty$ limit.

\subsubsection{Impact of heavy neutrinos}
\label{neutrino}

The effective interactions in Eq.~(\ref{LinINFTY}) involve five $\mathcal{N}_{i}$ coefficients, but this is partly due to the non-decoupling of neutrinos. To see this, let us add right handed neutrinos together with the Yukawa interaction $\bar{\nu}_{R}Y_{\nu}\ell_{L}\Phi_{1}$. Then, the effective amplitude in the linear representation is trivially obtained by extending the sum in Eq.~(\ref{LinAmpli}) to $f=u,d,e,\nu$. If, for the sake of the argument, we now take the infinite mass limit for all fermions, $m_{u,d,e,\nu}\rightarrow\infty$, the effective interactions in Eq.~(\ref{LinINFTY}) are recovered but with%
\begin{equation}
\mathcal{N}_{1}=\mathcal{N}_{2}=\frac{1}{12}\left(  N_{C}+1\right)  \left(
x+\frac{1}{x}\right)  \equiv\mathcal{N}_{\nu}\ .
\end{equation}
By contrast, it is amusing to remark that in the polar representation, the anomalous interactions of Eq.~(\ref{LJac}) are not modified by the presence of right-handed neutrinos since they are neutral under the whole SM gauge group. Yet, right-handed neutrinos do modify the derivative interactions in Eq.~(\ref{chiAV}) to $\chi_{V}^{\nu}=2\beta+x$ and $\chi_{A}^{\nu}=x$. The various triangle amplitudes are then obtained from those quoted in the previous section by replacing $\mathcal{N}_{1}\rightarrow\mathcal{N}_{\nu}$ and discarding the $\beta$ terms. While $\mathcal{M}(a^{0}\rightarrow ZZ,W^{+}W^{-})_{\text{\textrm{Der}}}^{AVV}$ now vanish in the $m_{u,d,e,\nu}\rightarrow\infty$ limit, this does not change our conclusions and $\mathcal{M}(a^{0}\rightarrow ZZ,W^{+}W^{-})_{\text{\textrm{Jac}}}$ remain different from $\mathcal{M}(A^{0}\rightarrow ZZ,W^{+}W^{-})_{\mathrm{Linear}}$ in that limit.

As a final note, it is interesting to remark that once certain that $\alpha$ and $\beta$ disappear from the physical amplitudes, one can chose to fix them as one wishes. In particular, as said before, they can be set to eliminate the electroweak $\theta$ term. Alternatively, one can set $\alpha=\beta=0$, thereby getting rid of the $a^{0}W_{\mu\nu}^{i}\tilde{W}^{i,\mu\nu}$ coupling altogether. As we have seen, this does not forbids $a^0\rightarrow W^+W^-$ since the amplitude is independent of $\alpha$ and $\beta$. Finally, this freedom can also be used to allow for the presence of a lepton number violating Majorana mass term, as well as an effective dimension-five $(\bar{\ell}_{L}^{C}H_{1})(\ell_{L}H_{1})/\Lambda$ operator. We will not explore such settings here, see Ref.~\cite{Latosinski:2012qj} for studies involving axions together with neutrino Majorana mass terms.

\subsection{Application to the DFSZ axion model}

The Peccei-Quinn model is ruled out experimentally because the coupling of the axion to fermions, tuned by the electroweak vacuum expectation value, is too large. One way to extend the model and render the axion invisible was proposed not long after the original idea and is called the DFSZ axion model~\cite{DFSZ}. Our goal here is to derive the electroweak couplings for that model.

The starting point is to extend the THDM by adding a complex scalar field $\phi$, singlet under the SM gauge symmetries, and constrain the scalar potential to%
\begin{align}
V_{\text{Scalar}} &  =m_{1}^{2}\Phi_{1}^{\dagger}\Phi_{1}+m_{2}^{2}\Phi
_{2}^{\dagger}\Phi_{2}+\frac{\lambda_{1}}{2}(\Phi_{1}^{\dagger}\Phi_{1}%
)^{2}+\frac{\lambda_{2}}{2}(\Phi_{2}^{\dagger}\Phi_{2})^{2}+\lambda_{3}%
(\Phi_{1}^{\dagger}\Phi_{1})(\Phi_{2}^{\dagger}\Phi_{2})+\lambda_{4}(\Phi
_{2}^{\dagger}\Phi_{1})(\Phi_{1}^{\dagger}\Phi_{2})\nonumber\\
&  +\mu^{2}\phi^{\dagger}\phi+\frac{\lambda}{2}(\phi^{\dagger}\phi)^{2}%
+a_{1}\phi^{\dagger}\phi\Phi_{1}^{\dagger}\Phi_{1}+a_{2}\phi^{\dagger}\phi
\Phi_{2}^{\dagger}\Phi_{2}+\left[  -\lambda_{12}^{2}\phi^{2}\Phi_{1}^{\dagger
}\Phi_{2}+h.c.\right]  \;.
\end{align}
In this way, the whole Lagrangian remains invariant under independent rephasing of the doublet provided the singlet has the charge:%
\begin{equation}
\Phi_{1}\rightarrow\exp(i\alpha_{1}\theta)\Phi_{1}\ ,\ \ \Phi_{2}%
\rightarrow\exp(i\alpha_{2}\theta)\Phi_{2}\ ,\ \ \phi\rightarrow\exp
(i\alpha\theta)\phi\ ,\ \ \alpha_{1}-\alpha_{2}=2\alpha\ .
\end{equation}
The Type II Yukawa couplings of Eq.~(\ref{YukQuark}) then imply chiral charges for the fermions when $\alpha_{1}\neq\alpha_{2}$, and part of the $U(1)_{1}\otimes U(1)_{2}$ symmetry is anomalous.

Let us take the linear representation (compare with Eq.~(\ref{HiggsBasis}))%
\begin{equation}
\Phi_{h}=\frac{1}{\sqrt{2}}\left(
\begin{array}[c]{c}%
G^{+}\\
v+\phi_{h}+iG^{0}%
\end{array}
\right)  \;,\;\;\Phi_{H}=\frac{1}{\sqrt{2}}\left(
\begin{array}[c]{c}%
\sqrt{2}H^{+}\\
\phi_{H}+i\phi_{A}%
\end{array}
\right)  \;,\;\;\phi=\frac{1}{\sqrt{2}}(v_{s}+\sigma_{S}+i\pi_{S})\ ,
\end{equation}
where $\Phi_{h,H}$ are given in terms of $\Phi_{1,2}$ in Eq.~(\ref{RotBeta}). Plugging these representations in the potential, and extracting the quadratic terms, the $G^{0}$ field is immediately massless and corresponds to the Would-be Goldstone of the $Z$ boson. On the other hand, the $\phi_{A}$ and $\pi_{S}$ fields mix,%
\begin{equation}
\left(
\begin{array}[c]{c}
A^{0}\\ \pi^{0}
\end{array}
\right)  =\frac{1}{\sqrt{v_{s}^{2}+v^{2}\sin^{2}2\beta}}\left(
\begin{array}[c]{cc}
v\sin2\beta & v_{s}\\
-v_{s} & v\sin2\beta
\end{array}
\right)  \left(
\begin{array}[c]{c}
\phi_{A}\\ \pi_{S}
\end{array}
\right)  \ ,
\end{equation}
with $M_{A^{0}}^{2}=0$ and $M_{\pi^{0}}^{2}=2\lambda_{12}(v^{2}\sin 2\beta+v_{s}^{2}/\sin2\beta)$. The $A^{0}$ field is the axion, and the $\pi^{0}$ field is a heavy pseudoscalar Higgs boson. While $A^{0}%
\rightarrow\phi_{A}$ when $v_{s}\rightarrow0$, reproducing the THDM discussed previously, the opposite holds when $v_{s}\gg v$~\cite{DFSZ}:
\begin{equation}
A^{0}=\pi_{S}+\frac{v}{v_{s}}\phi_{A}\sin2\beta+\mathcal{O}(v^{2}/v_{s}%
^{2})\ .
\end{equation}
Then, since $\pi_{S}$ is not coupled to fermions, it is easy to see that the $A^{0}\rightarrow V_{1}V_{2}$ amplitudes are simply rescaled by $v\sin 2\beta/v_{s}$, i.e.,%
\begin{align}
\mathcal{L}_{\mathrm{Linear}}^{\mathrm{eff,DFSZ}} &  =\frac{A^{0}\sin2\beta
}{16\pi^{2}v_{s}}\left(  g_{s}^{2}\mathcal{N}_{C}G_{\mu\nu}^{a}\tilde
{G}^{a,\mu\nu}+e^{2}\mathcal{N}_{em}F_{\mu\nu}\tilde{F}^{\mu\nu}+\frac{2e^{2}%
}{c_{W}s_{W}}\left(  \mathcal{N}_{0}-s_{W}^{2}\mathcal{N}_{em}\right)
Z_{\mu\nu}\tilde{F}^{\mu\nu}\right. \nonumber\\
&  \ \ \ \ \ \ \ \ \ \ \ \ \ \ \ \ \ \ \ \ \ \ \left.  +\frac{e^{2}}{c_{W}%
^{2}s_{W}^{2}}\left(  \mathcal{N}_{1}-2s_{W}^{2}\mathcal{N}_{0}+s_{W}%
^{4}\mathcal{N}_{em}\right)  Z_{\mu\nu}\tilde{Z}^{\mu\nu}+2\mathcal{N}%
_{2}g^{2}W_{\mu\nu}^{+}\tilde{W}^{-,\mu\nu}\right)  \ ,
\end{align}
with the same coefficients as in Eq.~(\ref{CoeffLin}). In particular, the axion couplings to $\gamma Z$, $ZZ$, and $W^{+}W^{-}$ are not aligned with the purely anomalous interactions one derives looking only at the $PQ$ charges of the fermion fields.

\subsection{Anomaly cancellation in generic axion models}

In the previous sections, we showed how the anomalies present in the non-linearly realized axion models necessarily cancel out, since they are spuriously generated by reparametrizing an anomaly-free model. Here, we want to follow the opposite route. Starting from a Non Renormalizable (NR) effective Lagrangian valid above the electroweak scale:%
\begin{align}
\mathcal{L}_{NR}^{\mathrm{eff}}  & =\frac{1}{2}\partial_{\mu}a^{0}%
\partial^{\mu}a^{0}-\frac{\partial_{\mu}a^{0}}{2v_{a}}\sum_{f=u,d,e,\nu
}\left(  \chi_{V}^{f}~\bar{\psi}_{f}\gamma^{\mu}\psi_{f}+\chi_{A}^{f}%
~\bar{\psi}_{f}\gamma^{\mu}\gamma_{5}\psi_{f}\frac{{}}{{}}\right) \nonumber\\
& +\frac{a^{0}}{16\pi^{2}v_{a}}\left(  g_{s}^{2}\mathcal{N}_{C}^{eff}G_{\mu
\nu}^{a}\tilde{G}^{a,\mu\nu}+g^{2}\mathcal{N}_{L}^{eff}W_{\mu\nu}\tilde
{W}^{\mu\nu}+g^{\prime2}\mathcal{N}_{Y}^{eff}B_{\mu\nu}\tilde{B}^{\mu\nu
}\right)  \ ,
\end{align}
where $v_{a}$ is some high scale, $\mathcal{N}_{C,L,Y}^{eff}$ as well as $\chi_{V,A}^{f}$ some a priori free Wilson coefficients, we want to show that whenever the anomalies cancel, this model collapses to a simple renormalizable theory.

To do this, starting from $\mathcal{L}_{NR}^{\mathrm{eff}}$, we require the anomalous interactions $\mathcal{N}_{C,L,Y}^{eff}$ to cancel with the local anomalous terms present in the $AVV$, $VAV$, and $AAA$ triangle graphs. Imposing this for the five channels $a^0\rightarrow V_{1}V_{2}$, $V_{1,2}=g,\gamma,Z,W$, four independent constraints are derived:%
\begin{align}
\mathcal{N}_{C}^{eff}  & =\frac{1}{2}(\chi_{A}^{u}+\chi_{A}^{d})\ ,\\
\mathcal{N}_{em}^{eff}  & \equiv\mathcal{N}_{L}^{eff}+\mathcal{N}_{Y}%
^{eff}=\frac{4}{9}N_{C}\chi_{A}^{u}+\frac{1}{9}N_{C}\chi_{A}^{d}+\chi_{A}%
^{e}\ ,\\
\mathcal{N}_{L}^{eff}  & =\frac{1}{6}N_{C}(\chi_{A}^{u}-\chi_{V}^{u})+\frac
{1}{12}N_{C}(\chi_{A}^{d}-\chi_{V}^{d})+\frac{1}{4}(\chi_{A}^{e}-\chi_{V}%
^{e})\ ,
\end{align}
together with%
\begin{equation}
\chi_{A}^{\nu}-\chi_{V}^{\nu}=\frac{1}{3}N_{C}(\chi_{A}^{u}-\chi_{V}%
^{u})-\frac{1}{3}N_{C}(\chi_{A}^{d}-\chi_{V}^{d})+\chi_{A}^{e}-\chi_{V}^{e}\ .
\end{equation}
Once these conditions are imposed, all the $a^0\rightarrow V_{1}V_{2}$ processes are entirely driven by the non-anomalous parts of the $AVV$ and $AAA$ triangle graphs (see Figs.~\ref{Fig3a},~\ref{Fig3b}, and \ref{Fig3c}). The $VAV$ triangles disappear since they have no mass-dependent terms. But then, these non-anomalous amplitudes satisfy the naive Ward identities, and are equivalently obtained from $PVV$ and $PAA$ triangles, provided%
\begin{equation}
\mathcal{L}_{R}^{\mathrm{eff}}=\frac{1}{2}\partial_{\mu}a^{0}\partial^{\mu
}a^{0}-i\sum_{f=u,d,e}\frac{m_{f}}{v_{a}}\chi_{P}^{f}\,a^{0}\bar{\psi}%
_{f}\gamma_{5}\psi_{f}\ ,\ \ \text{with }\chi_{P}^{f}=\chi_{A}^{f}%
\ .\label{DFSZlike}%
\end{equation}
Phenomenologically, $\mathcal{L}_{R}^{\mathrm{eff}}$ and $\mathcal{L}_{NR}^{\mathrm{eff}}$ are indistinguishable once anomaly cancellation is enforced, even if $\mathcal{L}_{R}^{\mathrm{eff}}$  has far fewer physical parameters. This can be understood as follows. Out of the initial ten parameters in $\mathcal{L}_{NR}^{\mathrm{eff}}$ for massless neutrino ($\chi_{V}^{\nu}=-\chi_{A}^{\nu}$), four are eliminated by the anomaly cancellation constraints. Then, the naive Ward identities are valid again. The three $\chi_{V}^{u,d,e}$ parameters disappear thanks to $\partial_{\mu}V^{\mu}=0$, and $\partial_{\mu}A^{\mu}=2imP$ leaves only the three $\chi_{P}^{f}=\chi_{A}^{f}$ parameters as truly physical.

Now, starting from $\mathcal{L}_{R}^{\mathrm{eff}}$, all the $a^0\rightarrow V_{1}V_{2}$ decay amplitudes are easily computed, and given by Eq.~(\ref{LinAmpli}). In the $m_{u,d,e}\rightarrow\infty$ limit, keeping neutrino massless, these amplitudes match onto the effective couplings in
Eq.~(\ref{LinINFTY}), with
\begin{subequations}
\begin{align}
\mathcal{N}_{C} &  =\frac{1}{2}\left(  \chi_{A}^{u}+\chi_{A}^{d}\right)  \ ,\\
\mathcal{N}_{em} &  =\frac{4}{9}N_{C}\chi_{A}^{u}+\frac{1}{9}N_{C}\chi_{A}%
^{d}+\chi_{A}^{e}\ ,\\
\mathcal{N}_{0} &  =\frac{1}{4}\left(  N_{C}\left(  \frac{2}{3}\chi_{A}%
^{u}+\frac{1}{3}\chi_{A}^{d}\right)  +\chi_{A}^{e}\right)  \ ,\\
\mathcal{N}_{1} &  =\frac{1}{12}\left(  N_{C}\left(  \chi_{A}^{u}+\chi_{A}%
^{d}\right)  +\chi_{A}^{e}\right)  \ ,\\
\mathcal{N}_{2} &  =\frac{1}{12}\left(  N_{C}\left(  \chi_{A}^{u}+\chi_{A}%
^{d}\right)  +\frac{3}{2}\chi_{A}^{e}\right)  \ .
\end{align}
\end{subequations}
In this limit, we thus do not recover the manifestly $SU(2)_{L}\otimes U(1)_{Y}$ invariant pattern of anomalous interactions of $\mathcal{L}_{NR}^{\mathrm{eff}}$. Finally, note that if one takes $\chi_{A}^{u}=\chi_{P}^{u}=x$, $\chi_{A}^{d}=\chi_{P}^{d}=\chi_{A}^{e}=\chi_{P}^{e}=1/x$, the results in Eq.~(\ref{CoeffLin}) are recovered. This shows that $\mathcal{L}_{R}^{\mathrm{eff}}$ is the most generic implementation of the PQ (or DFSZ) axion model, where the $U(1)_{PQ}$ charges of the SM fermions can take any value.

\section{Conclusion and perspective}
\label{Ccl}

In this paper, we derived the couplings of axions to gauge bosons. We started from a simplified toy model, and then extended the discussions to the KSVZ axion~\cite{KSVZ}, the original PQ axion~\cite{PQ}, and the DFSZ axion~\cite{DFSZ}. In each model, our strategy has been to match the axion decay modes computed using either a linear or a polar representation for the scalar field breaking the $U(1)_{PQ}$ symmetry. In this way, we were able to unravel the physics at play, and uncovered a number of interesting features:

\begin{itemize}
\item In the linear representation, the axion couplings to gauge bosons are not induced by the anomaly, but by non-anomalous pseudoscalar triangle graphs with fermions circulating in the loop. Though this has to our knowledge not been exploited before, these amplitudes can be identified with those of the pseudoscalar Higgs to gauge bosons calculated in the THDM, which have been known for a
long time~\cite{Gunion:1991cw}. In the $m_{u,d,e}\rightarrow \infty $ limit, they match onto the effective interactions 
\begin{align}
\mathcal{L}^{\mathrm{eff}}& =\frac{a^{0}}{16\pi ^{2}v}\left( g_{s}^{2}%
\mathcal{N}^{gg}G_{\mu \nu }^{a}\tilde{G}^{a,\mu \nu }+e^{2}\mathcal{N}%
^{\gamma \gamma }F_{\mu \nu }\tilde{F}^{\mu \nu }+\frac{2e^{2}}{c_{W}s_{W}}%
\left( \mathcal{N}_{1}^{\gamma Z}-s_{W}^{2}\mathcal{N}_{2}^{\gamma Z}\right)
Z_{\mu \nu }\tilde{F}^{\mu \nu }\right.   \notag \\
& \ \ \ \ \ \ \ \ \ \ \ \ \ \ \ \ \left. +\frac{e^{2}}{%
c_{W}^{2}s_{W}^{2}}\left( \mathcal{N}_{1}^{ZZ}-2s_{W}^{2}\mathcal{N}%
_{2}^{ZZ}+s_{W}^{4}\mathcal{N}_{3}^{ZZ}\right) Z_{\mu \nu }\tilde{Z}^{\mu
\nu }+2\mathcal{N}^{WW}g^{2}W_{\mu \nu }^{+}\tilde{W}^{-,\mu \nu }\right) \ ,
\label{EqCCL}
\end{align}%
with the coefficients shown in the first column of Table~\ref{TableCCL}. In the opposite limit $m_{u,d,e}\rightarrow 0$, all these amplitudes vanish since the axion couplings to fermions are proportional to the fermion masses.

\item In the context of axion models, it is customary to adopt a polar representation for the scalar fields. The pseudoscalar axion couplings to fermions are replaced by contact anomalous interactions to the gauge bosons and axion derivative interactions to the fermions. All these interactions are entirely fixed in terms of the assigned $PQ$ charges of the SM fermions. As shown in Table~\ref{TableCCL}, this parametrization is particularly convenient for $a^{0}\rightarrow \gamma \gamma $ and $a^{0}\rightarrow gg$ because the derivative interactions do not contribute in the $m_{u,d,e}\rightarrow \infty $ limit~\cite{Georgi:1986df,Bardeen:1986yb}, and thus the strengths of the $a^{0}\rightarrow \gamma \gamma $ and $a^{0}\rightarrow gg$ processes can be read off the anomalous couplings $a^{0}F_{\mu \nu }\tilde{F}^{\mu \nu }$ and $a^{0}G_{\mu \nu }^{a}\tilde{G}^{a,\mu \nu }$. By contrast, and contrary to what is conjectured in Ref.~\cite{Georgi:1986df}, the central result of this paper is that for chiral gauge theories in which chiral fermions have non-trivial $PQ$ charges, the strengths of the $a^{0}\rightarrow \gamma Z$, $ZZ$, and $W^{+}W^{-}$ processes do not match the anomalous couplings, even in the $m_{u,d,e}\rightarrow \infty $ limit, as evident comparing the first and second columns of Table~\ref{TableCCL}. They thus cannot be encoded into the $SU(2)_{L}\otimes U(1)_{Y}$ invariant effective interactions $a^{0}B_{\mu\nu }\tilde{B}^{\mu \nu }\ $and $a^{0}W_{\mu \nu }^{i}\tilde{W}^{i,\mu \nu }$.

\begin{table}[t] \centering
\begin{tabular}{c|c|cc|c}
\hline
Linear & \multicolumn{4}{|c}{Polar} \\ 
\multicolumn{1}{c|}{$a^0\bar{\psi}\gamma _{5}\psi $} & Anomalous & 
\multicolumn{2}{|c|}{$\partial _{\mu }a^0\bar{\psi}\gamma ^{\mu }\gamma
_{5}\psi $} & $\partial _{\mu }a^0\bar{\psi}\gamma ^{\mu }\psi $ \\ 
\multicolumn{1}{c|}{} & interactions & $AVV$ & $AAA$ & $VAV$ \\ \hline
\multicolumn{1}{l|}{$\mathcal{N}^{gg}=\frac{1}{2}\left( x+\frac{1}{x}\right) 
$} & $\mathcal{N}^{gg}$ & 0 & $-$ & $-\rule[-0.1in]{0in}{0.3in}$ \\ \hline
\multicolumn{1}{l|}{$\mathcal{N}^{\gamma \gamma }=\frac{4}{3}\left( x+\frac{1%
}{x}\right) $} & $\mathcal{N}^{\gamma \gamma }$ & 0 & $-$ & $-%
\rule[-0.1in]{0in}{0.3in}$ \\ \hline
\multicolumn{1}{l|}{$\mathcal{N}_{1}^{\gamma Z}=\frac{1}{2}\left( x+\frac{1}{%
x}\right) $} & $\mathcal{N}_{L}$ & 0 & $-$ & $\mathcal{N}_{1}^{\gamma Z}-%
\mathcal{N}_{L}\rule[-0.1in]{0in}{0.3in}$ \\ 
\multicolumn{1}{l|}{$\mathcal{N}_{2}^{\gamma Z}=\mathcal{N}^{\gamma \gamma }$%
} & $\mathcal{N}^{\gamma \gamma }$ & 0 & $-$ & 0$\rule[-0.1in]{0in}{0.3in}$
\\ \hline
\multicolumn{1}{l|}{$\mathcal{N}_{1}^{ZZ}=\frac{1}{4}x+\frac{1}{3x}$} & $%
\mathcal{N}_{L}$ & $\frac{\beta }{16}\ $ & $-\frac{1}{2}\mathcal{N}_{1}^{ZZ}+%
\frac{\beta }{16}$ & $\frac{3}{2}\mathcal{N}_{1}^{ZZ}-\mathcal{N}_{L}-\frac{%
\beta }{8}\rule[-0.1in]{0in}{0.3in}$ \\ 
\multicolumn{1}{l|}{$\mathcal{N}_{2}^{ZZ}=\mathcal{N}_{1}^{\gamma Z}$} & $%
\mathcal{N}_{L}$ & 0 & 0 & $\mathcal{N}_{2}^{ZZ}-\mathcal{N}_{L}%
\rule[-0.1in]{0in}{0.3in}$ \\ 
\multicolumn{1}{l|}{$\mathcal{N}_{3}^{ZZ}=\mathcal{N}^{\gamma \gamma }$} & $%
\mathcal{N}^{\gamma \gamma }$ & 0 & 0 & 0$\rule[-0.1in]{0in}{0.3in}$ \\ 
\hline
\multicolumn{1}{l|}{$\mathcal{N}^{WW}=\frac{x}{4}+\frac{3}{8x}$} & $\mathcal{%
N}_{L}$ & $\frac{3}{2}\mathcal{N}^{WW}-\frac{3}{2}\mathcal{N}_{1}^{\gamma Z}+%
\frac{\beta }{16}$ & $-\frac{1}{2}\mathcal{N}^{WW}+\frac{\beta }{16}$ & $%
\frac{3}{2}\mathcal{N}_{1}^{\gamma Z}-\mathcal{N}_{L}-\frac{\beta }{8}%
\rule[-0.1in]{0in}{0.3in}$ \\ \hline
\end{tabular}%
\caption
{Coefficients of the effective axion to gauge boson couplings of Eq.~(\ref{EqCCL}) in the $m_{u,d,e} \rightarrow \infty$ limit. For the linear representation, those are found directly from the THDM amplitudes of Ref.~\cite{Gunion:1991cw}. For the polar representation, the contributions of the local anomalous terms and that of the triangle amplitudes built on the axial ($A$) and vector ($V$) derivative interactions have to be added together. The fact that only the three independent coefficients $\mathcal{N}^{gg}$, $\mathcal{N}^{\gamma\gamma}$, and $\mathcal{N}_L$ occur for the local anomalous interactions comes from their $SU(2)_L \otimes U(1)_Y$ invariance. The coefficient $\mathcal{N}_L = -1/2 (3 \alpha + \beta)$, with $\alpha$ and $\beta$ being the free parameters tuning the $U(1)_{\mathcal{B}}$ and $U(1)_{\mathcal{L}}$ components of $U(1)_{PQ}$. The explicit presence of $\beta$ in the $a^0 \rightarrow ZZ, WW$ triangles is due to the peculiar nature of the neutrinos, which are kept massless. They do not contribute in the linear representation but have to explicitly appear in the anomalous interactions and derivative terms since those are $SU(2)_L \otimes U(1)_Y$ invariant.}%
\label{TableCCL}%
\end{table}%

\item The main reason for this mismatch is due to the presence of triangle graphs arising from the derivative interactions $\partial _{\mu }a^{0}\bar{\psi}\gamma ^{\mu }\gamma _{5}\psi $ and $\partial _{\mu }a^{0}\bar{\psi}\gamma ^{\mu }\psi $, which do not vanish in the $m_{u,d,e}\rightarrow \infty $ limit for chiral gauge theories. Importantly, even the vector couplings play a role since the anomalous breaking of the axionic vector current conservation enters through the $VAV$ triangle graphs (see Figs.~\ref{Fig3b} and \ref{Fig3c}). One important result of this paper is the proof that once all these triangle contributions (last three columns of Table~\ref{TableCCL}) are summed with the local anomalous amplitudes (second column of Table~\ref{TableCCL}), the THDM results are recovered. Without surprise, physical observables do not depend on the chosen parametrization, and this further confirms that the THDM calculation is correct.

\item Our analysis also proves that the axion to gauge boson processes are not induced by the anomaly. This is obvious in the linear representation since the fermion loops driving these processes are not anomalous. In the polar representation, what happens is that the local anomalous interactions precisely cancel with the anomalies present in triangle graphs induced by the vector and axial derivative interactions, leaving the non-anomalous pseudoscalar triangles of the THDM as the only surviving mechanism driving all the axion to gauge bosons processes. Ultimately, this precise cancellation is ensured by the anomalous vector and axial Ward identities. It is required because all the amplitudes must vanish in the $m_{u,d,e,\nu }\rightarrow 0$ limit, but the anomalous contributions are independent of the fermion mass\footnote{This interpretation is compatible with the matching observed in Table~\ref{TableCCL} in the $m_{u,d,e}\rightarrow \infty $ limit for $a^{0}\rightarrow \gamma \gamma ,gg$. Indeed, in the polar representation, these amplitudes receive two contributions, $\mathcal{M}_{ano}$ from the local anomalous term, and $\mathcal{M}_{der}$ from the triangle amplitudes built from $\partial _{\mu }a^0\bar{\psi}\gamma ^{\mu }\gamma
_{5}\psi$. But in this case, the axial Ward identity translates as $\mathcal{M}_{der} = \mathcal{M}_{lin}  - \mathcal{M}_{ano}$, with $\mathcal{M}_{lin}$ the THDM amplitude. So, even if parametrically, $\mathcal{M}_{lin} = \mathcal{M}_{ano}$ and $\mathcal{M}_{der}=0$ when $m_{u,d,e}\rightarrow \infty $, $a^{0}\rightarrow \gamma \gamma ,gg$ are not induced by the anomaly since it cancels out in $\mathcal{M}_{ano}+\mathcal{M}_{der}$.}.

\item In all axion models where chiral fermions have $PQ$ charges but conserve baryon and lepton numbers, both the vector couplings $\partial_{\mu }a^{0}\bar{\psi}\gamma ^{\mu }\psi $ and the $a^{0}W_{\mu \nu }^{i}\tilde{W}^{i,\mu \nu }$, $a^{0}B_{\mu \nu }\tilde{B}^{\mu \nu }$ anomalous contact interactions depend on some free parameters~\cite{Georgi:1986df}. In Table~\ref{TableCCL}, these free parameters enter into $\mathcal{N}_{L}$, and are related to the freedom to choose the $PQ$ charge of left-handed fermions (there are also explicit occurrence of the free parameter $\beta $, but this is an artifact of keeping the neutrino massless). It is only once the failure of the naive vector Ward identity in the $VAV$ triangle graphs is properly accounted for that these free parameters cancel out, as they should~\cite{Dreiner:2008tw}. To understand how this cancellation is connected with the conservation of baryon and lepton numbers $\mathcal{B}$ and $\mathcal{L}$, notice first that the freedom to choose the $PQ$ charges of the fermions is actually due to the invariance under $U(1)_{\mathcal{B}}$ and $U(1)_{\mathcal{L}}$. In general, the $U(1)_{PQ}$ symmetry has two arbitrary components aligned with $U(1)_{\mathcal{B}}$ and $U(1)_{\mathcal{L}}$. But then, these components translate into derivative couplings of the axion to the $\mathcal{B}$ and $\mathcal{L}$ fermionic vectorial currents together with a local $a^{0}W_{\mu \nu }^{i}\tilde{W}^{i,\mu \nu }$ coupling from the $\mathcal{B}+\mathcal{L}$ anomaly. Clearly, all these couplings are spurious when the axion is a Goldstone boson living in a $\mathcal{B}$ and $\mathcal{L}$-invariant vacuum. Since $\mathcal{B}$ and $\mathcal{L}$ are not spontaneously broken, $\mathcal{N}_{L}$ must systematically drop out of physical observables, and Table~\ref{TableCCL} shows that this indeed occurs.

\item In the linear representation, the calculation is done in the electroweak broken phase throughout. In the $m_{u,d,e}\rightarrow \infty $ limit, the amplitudes do not match onto $SU(2)_{L}\otimes U(1)_{Y}$ invariant operators, as shown in Table~\ref{TableCCL}. Though this is somewhat expected, what is not obvious is the very peculiar way in which the $SU(2)_{L}\otimes U(1)_{Y}$ symmetry breaking seeps in. Indeed, in the non-linear representation, the anomalous interactions and the derivative interactions are manifestly $SU(2)_{L}\otimes U(1)_{Y}$ invariant once the SM fermions are assigned $PQ$ charges in an $SU(2)_{L}\otimes U(1)_{Y}$ invariant way. Further, evidently, the anomalies present in the triangle graphs built on the derivative interactions do not break $SU(2)_{L}\otimes U(1)_{Y}$ since that would put the SM gauge invariance itself in jeopardy. So, in that representation, the fact that the chiral $SU(2)_{L}\otimes U(1)_{Y}$ symmetry is spontaneously broken enters only and entirely through the fermion masses. The triangle graph explicitly break the $SU(2)_{L}\otimes U(1)_{Y}$ symmetry because the $m_{u,d,e}\rightarrow \infty $ limit is incompatible with that symmetry. In the opposite limit $m_{u,d,e}\rightarrow 0$, the $SU(2)_{L}\otimes U(1)_{Y}$ symmetry is recovered but in a very trivial way: the triangle graphs precisely cancel the local anomalous term and all the amplitudes simply vanish.

\item We derived generalized forms for all the triangle anomalies, including the terms proportional to the fermion masses, see Eqs.~(\ref{AnoAVVm}) and~(\ref{AnoAAAm}). This requires going beyond usual regularization procedures. To our knowledge, and except those based on dispersion relations~\cite{Frishman:1980dq}, only that described by Weinberg in Ref.~\cite{Weinberg:1996kr} offers sufficient freedom to choose which current in the $AVV$ and $AAA$ triangles is to carry the anomaly. This is crucial to deal with chiral gauge theories, in which some axial currents have to remain
anomaly-free to preserve gauge invariance, as well as for treating the $\mathcal{B}$ and $\mathcal{L}$ vectorial currents~\cite{Dreiner:2008tw}.
\end{itemize}

In the present paper, we consider only true axion models, those designed to solve the strong $CP$ puzzle and for which the axion ends up extremely light and extremely weakly coupled to SM fields. However, our results could have consequences for so-called Axion-Like Particle (ALP) searches (see e.g. Refs.~\cite{Salvio:2013iaa,Jaeckel:2015jla,Bauer:2017ris,Ebadi:2019gij,Brivio:2017ije,Gavela:2019wzg,Bauer:2018uxu}). Specifically, our analysis shows that an adequate effective parametrization is
\begin{equation}
\mathcal{L}_{ALP}=\frac{1}{2}(\partial_{\mu}a^{0}\partial^{\mu}a^{0}-m_{a}%
^{2}a^{0}a^{0})+\mathcal{L}_{\mathrm{DFSZ-like}}+\mathcal{L}%
_{\mathrm{KSVZ-like}}\ ,
\end{equation}
with from Eq.~(\ref{DFSZlike}),%
\begin{equation}
\mathcal{L}_{\mathrm{DFSZ-like}}=-i\sum_{f=u,d,e}\frac{m_{f}}{v_{a}}%
\chi_{P}^{f}\,a^{0}\bar{\psi}_{f}\gamma_{5}\psi_{f}\ ,
\end{equation}
for some free parameters $\chi_{P}^{f}$, and from Eq.~(\ref{KSVZlike})
\begin{equation}
\mathcal{L}_{\mathrm{KSVZ-like}}=\frac{a^{0}}{16\pi^{2}v_{a}}\left(  g_{s}%
^{2}\mathcal{N}_{C}^{eff}G_{\mu\nu}^{a}\tilde{G}^{a,\mu\nu}+g^{2}%
\mathcal{N}_{L}^{eff}W_{\mu\nu}^{i}\tilde{W}^{i,\mu\nu}+g^{\prime2}%
\mathcal{N}_{Y}^{eff}B_{\mu\nu}\tilde{B}^{\mu\nu}\right)  \ ,
\end{equation}
for some free parameters $\mathcal{N}_{C,L,Y}^{eff}$. Typically, $\mathcal{L}_{\mathrm{KSVZ-like}}$ encodes the effects of heavy vector-like fermions, while $\mathcal{L}_{\mathrm{DFSZ-like}}$ encode the anomaly-free impact of charging the SM fermions under some global $U(1)_{PQ}$ symmetry whose current is coupled to the $a^{0}$ field. The impact of some new heavy chiral fermions (like a sequential fourth generation or a heavy neutrino) is not included, but could be by extending the sum in $\mathcal{L}_{\mathrm{DFSZ-like}}$ to those states also. This parametrization of $\mathcal{L}_{\mathrm{DFSZ-like}}$ ensures $a^{0}$ is truly axion-like: no $\mathcal{B}$ or $\mathcal{L}$ violating effects are implicit, only physical free parameters are introduced, and UV divergences should be under control. Still, it is important to stress that the $a^{0}\rightarrow\gamma Z$, $ZZ$, and $W^{+}W^{-}$ amplitudes induced by $\mathcal{L}_{\mathrm{DFSZ-like}}$ are entirely tuned by triangle graphs, and in the $m_{f}\rightarrow\infty$ limit, they do not match the pattern of $\mathcal{L}_{\mathrm{KSVZ-like}}$ but rather that in Eq.~(\ref{LinINFTY}).


\begin{thebibliography}{99}                                                                                               %
\bibitem {PQ}R.~D.~Peccei and H.~R.~Quinn,
Phys.\ Rev.\ Lett.\ \textbf{38} (1977) 1440;
R.~D.~Peccei and H.~R.~Quinn,
Phys.\ Rev.\ D \textbf{16} (1977) 1791.

\bibitem {Weinberg:1977ma}S.~Weinberg,
Phys.\ Rev.\ Lett.\ \textbf{40} (1978) 223.

\bibitem {Wilczek:1977pj}F.~Wilczek,
Phys.\ Rev.\ Lett.\ \textbf{40} (1978) 279.

\bibitem{Shtabovenko:2016sxi}
  V.~Shtabovenko, R.~Mertig and F.~Orellana,
  Comput.\ Phys.\ Commun.\  {\bf 207} (2016) 432
  [arXiv:1601.01167 [hep-ph]].
  
\bibitem {diCortona:2015ldu}G.~Grilli di Cortona, E.~Hardy, J.~Pardo Vega and
G.~Villadoro,
JHEP \textbf{1601} (2016) 034 [arXiv:1511.02867 [hep-ph]].

\bibitem {Bardeen:1977bd}W.~A.~Bardeen and S.-H.~H.~Tye,
Phys.\ Lett.\ \textbf{74B} (1978) 229.

\bibitem {Kim:1986ax}J.~E.~Kim,
Phys.\ Rept.\ \textbf{150} (1987) 1.

\bibitem{Marsh:2015xka}
  D.~J.~E.~Marsh,
  Phys.\ Rept.\  {\bf 643} (2016) 1
  [arXiv:1510.07633 [astro-ph.CO]].

\bibitem {Georgi:1986df}H.~Georgi, D.~B.~Kaplan and L.~Randall,
Phys.\ Lett.\ \textbf{169B} (1986) 73.

\bibitem {Bardeen:1986yb}W.~A.~Bardeen, R.~D.~Peccei and T.~Yanagida,
Nucl.\ Phys.\ B \textbf{279} (1987) 401.

\bibitem {Alonso-Alvarez:2018irt}G.~Alonso-\'{A}lvarez, M.~B.~Gavela and
P.~Quilez,
arXiv:1811.05466 [hep-ph].

\bibitem {KSVZ}J.~E.~Kim,
Phys.\ Rev.\ Lett.\ \textbf{43} (1979) 103;
M.~A.~Shifman, A.~I.~Vainshtein and V.~I.~Zakharov,
Nucl.\ Phys.\ B \textbf{166} (1980) 493.

\bibitem {Adler:1969er}S.~L.~Adler and W.~A.~Bardeen,
Phys.\ Rev.\ \textbf{182} (1969) 1517.

\bibitem {Ametller:1983ec}L.~Ametller, L.~Bergstrom, A.~Bramon and E.~Masso,
Nucl.\ Phys.\ B \textbf{228} (1983) 301.

\bibitem {Weinberg:1996kr}S.~Weinberg, \textit{The quantum theory of fields.
Vol. 2: Modern applications}, Cambridge, 1996.

\bibitem {Bilal:2008qx}A.~Bilal, \textit{Lectures on Anomalies},
arXiv:0802.0634 [hep-th].

\bibitem {Frishman:1980dq}Y.~Frishman, A.~Schwimmer, T.~Banks and
S.~Yankielowicz,
Nucl.\ Phys.\ B \textbf{177} (1981) 157.

\bibitem {Gunion:1989we}J.~F.~Gunion, H.~E.~Haber, G.~L.~Kane and S.~Dawson,
\textit{The Higgs Hunter's Guide}, Front.\ Phys.\ \textbf{80} (2000) 1.

\bibitem {Gunion:1991cw}J.~F.~Gunion, H.~E.~Haber and C.~Kao,
Phys.\ Rev.\ D \textbf{46} (1992) 2907.

\bibitem {Djouadi:2005gi}A.~Djouadi,
Phys.\ Rept.\ \textbf{457} (2008) 1 [hep-ph/0503172].

\bibitem {Kamefuchi:1961sb}S.~Kamefuchi, L.~O'Raifeartaigh and A.~Salam,
Nucl.\ Phys.\ \textbf{28} (1961) 529.

\bibitem {Dreiner:2008tw}H.~K.~Dreiner, H.~E.~Haber and S.~P.~Martin,
Phys.\ Rept.\ \textbf{494} (2010) 1 [arXiv:0812.1594 [hep-ph]].

\bibitem {Latosinski:2012qj}A.~Latosinski, K.~A.~Meissner and H.~Nicolai,
Nucl.\ Phys.\ B \textbf{868} (2013) 596 [arXiv:1203.3886 [hep-ph]].

\bibitem {DFSZ}M.~Dine, W.~Fischler and M.~Srednicki,
Phys.\ Lett.\ \textbf{104B} (1981) 199;
A.~R.~Zhitnitsky,
Sov.\ J.\ Nucl.\ Phys.\ \textbf{31} (1980) 260 [Yad.\ Fiz.\ \textbf{31} (1980)
497].

\bibitem {Salvio:2013iaa}A.~Salvio, A.~Strumia and W.~Xue,
JCAP \textbf{1401} (2014) 011 [arXiv:1310.6982 [hep-ph]].

\bibitem {Jaeckel:2015jla}J.~Jaeckel and M.~Spannowsky,
Phys.\ Lett.\ B \textbf{753} (2016) 482 [arXiv:1509.00476 [hep-ph]].

\bibitem {Bauer:2017ris}M.~Bauer, M.~Neubert and A.~Thamm,
JHEP \textbf{1712} (2017) 044 [arXiv:1708.00443 [hep-ph]].

\bibitem {Ebadi:2019gij}J.~Ebadi, S.~Khatibi and M.~Mohammadi Najafabadi,
arXiv:1901.03061 [hep-ph].

\bibitem {Brivio:2017ije}I.~Brivio, M.~B.~Gavela, L.~Merlo, K.~Mimasu,
J.~M.~No, R.~del Rey and V.~Sanz,
Eur.\ Phys.\ J.\ C \textbf{77} (2017) no.8, 572 [arXiv:1701.05379 [hep-ph]].

\bibitem {Gavela:2019wzg}M.~B.~Gavela, R.~Houtz, P.~Quilez, R.~Del Rey and
O.~Sumensari,
arXiv:1901.02031 [hep-ph].

\bibitem {Bauer:2018uxu}M.~Bauer, M.~Heiles, M.~Neubert and A.~Thamm,
Eur.\ Phys.\ J.\ C \textbf{79} (2019) no.1, 74 [arXiv:1808.10323 [hep-ph]].

\end{thebibliography}
\end{document}